\documentclass[journal=jpclcd,manuscript=letter]{achemso}
\usepackage{amsmath,amssymb}
\usepackage{float}
\usepackage{enumerate}
\usepackage[version=3]{mhchem} 
\usepackage{array,longtable}
\usepackage{color} 

\author{Deepika Gill}
\author{Preeti Bhumla}
\author{Manish Kumar}
\author{Saswata Bhattacharya}
\email{saswata@physics.iitd.ac.in [SB]}
\phone{+91-11-2659 1359}
\fax{+91-11-2658 2037}
\affiliation[Indian Institute of Technology Delhi]
{Department of Physics, Indian Institute of Technology Delhi, New Delhi, India}
\title[An \textsf{achemso} demo]
  {High-Throughput Screening for Band gap Engineering by Sublattice Mixing of Cs$_2$AgBiCl$_6$ from First-Principles}
\begin{document}
\begin{abstract}
\begin{tocentry}
	\begin{figure}[H]
		\includegraphics[width=1.0\columnwidth,clip]{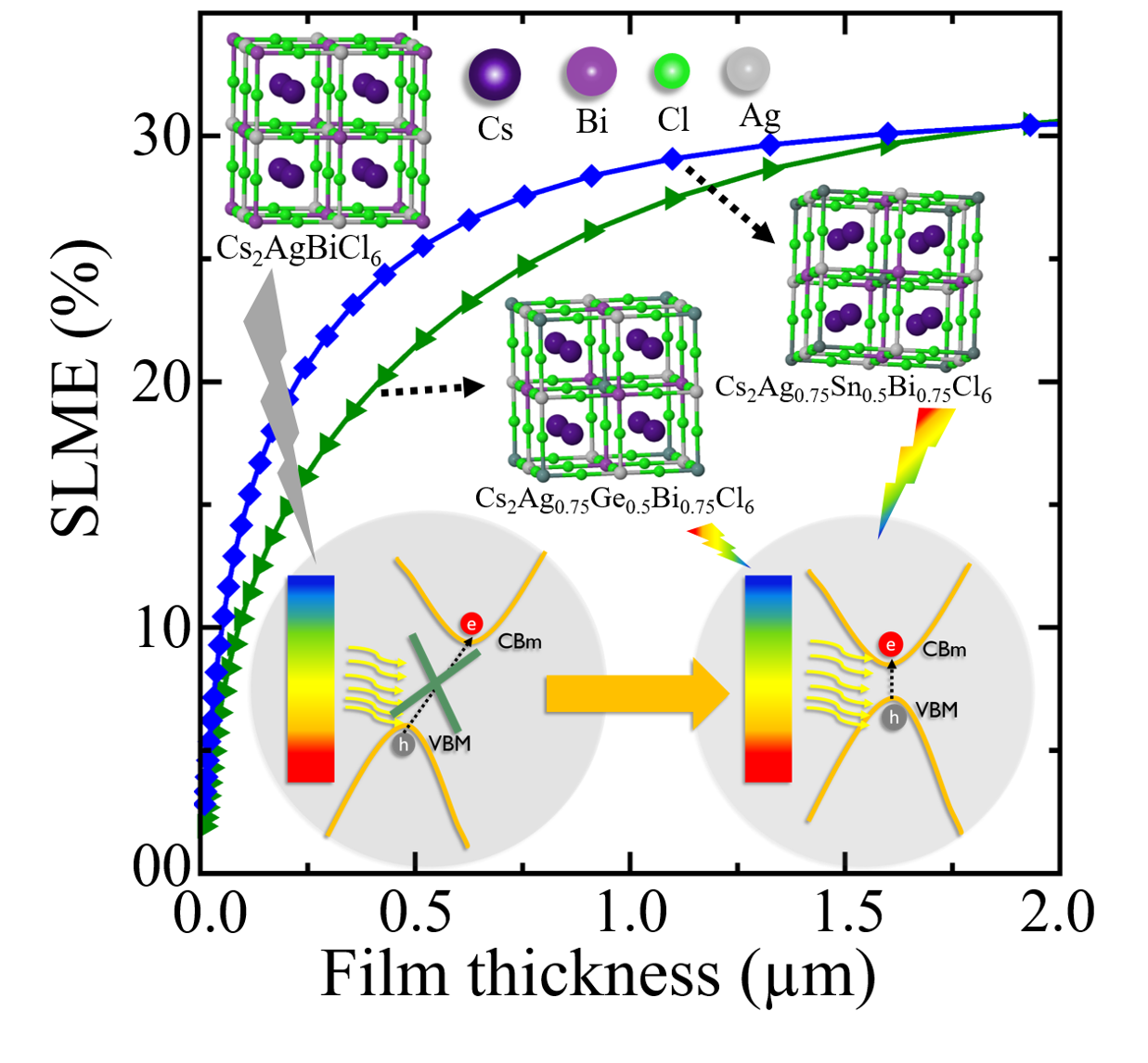}
	\end{figure}
\end{tocentry}
The lead-free double perovskite material (viz. Cs$_2$AgBiCl$_6$) has emerged as an efficient and environmentally friendly alternative to lead halide perovskites. To make Cs$_2$AgBiCl$_6$ optically active in the visible region of solar spectrum, band gap engineering approach has been undertaken. Using Cs$_2$AgBiCl$_6$ as a host, band gap and optical properties of Cs$_2$AgBiCl$_6$ have been modulated by alloying with M(I), M(II), and M(III) cations at Ag-/Bi-sites. Here, we have employed density functional theory (DFT) with suitable exchange-correlation functionals in light of spin-orbit coupling (SOC) to determine the stability, band gap and optical properties of different compositions, that are obtained on Ag-Cl and Bi-Cl sublattices mixing. On analyzing the 64 combinations within Cs$_2$AgBiCl$_6$, we have identified 19 promising configurations having band gap sensitive to solar cell applications. The most suitable configurations with Ge(II) and Sn(II) substitutions have spectroscopic limited maximum efficiency (SLME) of 32.08\% and 30.91\%, respectively, which are apt for solar cell absorber.
\end{abstract}
Inorganic-organic (IO) hybrid halide perovskites (AM(\textrm{II})X$_3$, A = methylammonium (MA$^+$), formamidinium (FA$^+$); M(\textrm{II}) = Pb$^{2+}$; X = Cl$^{-}$, Br$^{-}$, I$^{-}$) have brought a huge revolution in the field of photovoltaics~\cite{frost2014atomistic, de2016guanidinium, zhang2018planar}. These alluring materials exhibit high absorption coefficient, long carrier diffusion length, high carrier mobility, low trap density, low manufacturing cost and high defect tolerance~\cite{r1,doi:10.1021/jz500858a,PhysRevB.101.054108,doi:10.1021/acsenergylett.6b00196}. Starting from 3.8\%, their power conversion efficiency (PCE) has risen to 22.7\%~\cite{r1, r2, r3} in just a decade. Despite these attainments, the intrinsic instability owing to monovalent organic cation~\cite{park2019intrinsic} and toxicity of Pb~\cite{babayigit2016toxicity} hinder their large scale commercialization. To address these problems, extensive efforts have been paid to find stable and green alternatives for optoelectronic applications. In the recent years, despite large success of IO hybrid perovskites, researchers are getting back to inorganic perovskites as the former suffers from intrinsic stability. It has been reported that the problem of intrinsic instability can be overcome by replacing organic cation with inorganic cation i.e., Cs$^{+}$. This not only upgrades the thermal stability but also exceeds device's life span~\cite{kulbak2015important, liang2016all}. However, in CsPbX$_3$, toxic nature of Pb is still a big issue and replacement of Pb with some non-toxic element provides a permanent solution. Also, complete replacement of Pb with elements belonging to the same group like Sn and Ge is not suitable due to their tendency of oxidization from +2 state to +4 state~\cite{S1, S2}. Considering any other divalent cation in place of Pb results in poor optoelectronic properties owing to their large band gap~\cite{S3, S4}. Alternatively, without varying the total number of valence electrons, two Pb$^{2+}$ cations can be transmuted by one monovalent (M(I)) and one trivalent (M(III)) cation. It leads to a new configuration i.e., Cs$_{2}$M(I)M(III)X$_{6}$~\cite{cai2019high,chen2019yb}, which procures a double perovskite structure. Following this kind of design strategy, a few experimentally reported double perovskites are Cs$_{2}$AgBiX$_{6}$ [X=Cl, Br, I]~\cite{S5, S6, S7}, Cs$_{2}$AgSbX$_{6}$~\cite{S8, S9}, Cs$_{2}$AgInX$_{6}$~\cite{S10,locardi2018colloidal} and Cs$_{2}$InM(III)X$_{6}$ [M(III)=Sb, Bi]~\cite{S11}. However, none of them are ideal for solar cell applications due to the imperfections associated with them. In case of Cs$_{2}$AgM(III)X$_{6}$ [M(III)=Sb,Bi], the large indirect band gap results in poor solar absorption. On the other hand, parity forbidden transition and inevitable conversion of In$^{1+}$ to In$^{3+}$ in Cs$_{2}$AgInX$_{6}$~\cite{S10} and Cs$_{2}$InM(III)X$_{6}$ [M(III)=Sb,Bi]~\cite{S13}, respectively, lead to degradation of photovoltaic performance.

In this Letter,
we present an intensive theoretical study on band gap transmutation of Cs$_{2}$AgBiCl$_{6}$ by means of sublattice mixing. The sublattice mixing is done by substituting M(I) at Ag-sites, M(II) at Ag- and Bi-sites simultaneously, and M(III) at Bi-sites in various concentrations for enhancing the optical properties of Cs$_{2}$AgBiCl$_{6}$. A high-throughput screening is performed by carrying out the hierarchical computations employing state-of-the-art first-principles based methodologies under the framework of density functional theory (DFT). We start doing a lot of pre-screening of a large number of configurations with DFT using generalized gradient approximation based exchange-correlation ($\epsilon_{xc}$) functional (PBE~\cite{perdew1992atoms}) and following that the promising candidate structures are further analyzed using hybrid DFT with HSE06~\cite{heyd2003hybrid}. The latter $\epsilon_{xc}$ functional helps for more accurate understanding of the excited state properties. Note that in all the above calculations (viz. PBE or HSE06), the effect of spin-orbit coupling (SOC) is always taken into consideration. This is a crucial step to determine the accurate band gap and band-edge positions of these systems due to presence of heavy metal atoms.
We have started with 64 sets of different combinations of metals M(I), M(II), and M(III)  respectively. Firstly, the structural stability is predicted using the Goldschmidt's tolerance factor and octahedral factor. It is worth noting that structural stability alone is not sufficient for the formation of perovskites. Hence, to validate the material's thermodynamic stability, the enthalpy of decomposition ($\Delta$H$_{\textrm{D}}$) is calculated. Then, from $\Delta$H$_{\textrm{D}}$ and band gap range (which expands the spectral response in visible region), the promising stable double perovskite configurations are identified.
Following identification of such potential candidate structures, detailed electronic structure is carried out alongside of computing optical properties. The real and imaginary parts of the dielectric function are analyzed to understand the effect of sublattice mixing in Cs$_{2}$AgBiCl$_{6}$ for transmutation of band gap. Subsequently, we determine spectroscopic limited maximum efficiency (SLME) of all the stable configurations that possess direct band gap, to determine efficient solar cell absorber. 


Initially, the benchmarking of $\epsilon_{xc}$ functional has been performed by calculating the band gap of pristine system viz. Cs$_2$AgBiCl$_6$. 
The band gap of Cs$_2$AgBiCl$_6$ with PBE $\epsilon_{xc}$ functional is 2.06 eV, which is not in agreement with the experimental value of 2.77 eV~\cite{mcclure2016cs2agbix6}. As this system contains heavy metal atom (viz. Bi), the inclusion of SOC becomes important~\cite{PhysRevB.101.054108}.
However, incorporation of SOC with PBE underestimates the band gap (1.68 eV) further due to splitting of the conduction band minimum (CBm). The latter gets shifted to lower energy toward Fermi-level, whereas the valence band maximum (VBM) remains unaffected (see Figure S1a in Supplementary Information (SI)). 
On the other hand, to include the self-interaction error of a many-electron system in the expression of $\epsilon_{xc}$ functional, advanced hybrid $\epsilon_{xc}$ functional viz. HSE06 becomes essential. It gives a band gap of 3.15 eV (without SOC, HSE06 only) and 2.60 eV (with SOC, HSE06+SOC) respectively with default (0.25) Hartree-Fock exchange fraction ($\alpha$) (see SI Figure S2). On increasing $\alpha$ to $0.30$ and $0.35$, we have obtained band gaps of $2.79$ and $2.99$ eV, respectively using HSE06+SOC (see Figure S1 and S2 in SI). Note that using HSE06+SOC w.r.t the experimental value, though the band gap with $\alpha$ = $0.30$ is more accurate than that of default value (i.e. $\alpha$ = $0.25$), we still have proceeded with default one for further calculations. This is due to the fact that on alloying with various metals, the value of $\alpha$ can vary from one system to other, and determining the same without experimental results is next to impossible for new configurations. Therefore, we expect, atleast the default $\alpha$ should give a correct trend qualitatively even if the actual numbers may differ marginally as in the case of pristine Cs$_2$AgBiCl$_6$.
\begin{figure}[H]
	\includegraphics[width=0.8\columnwidth,clip]{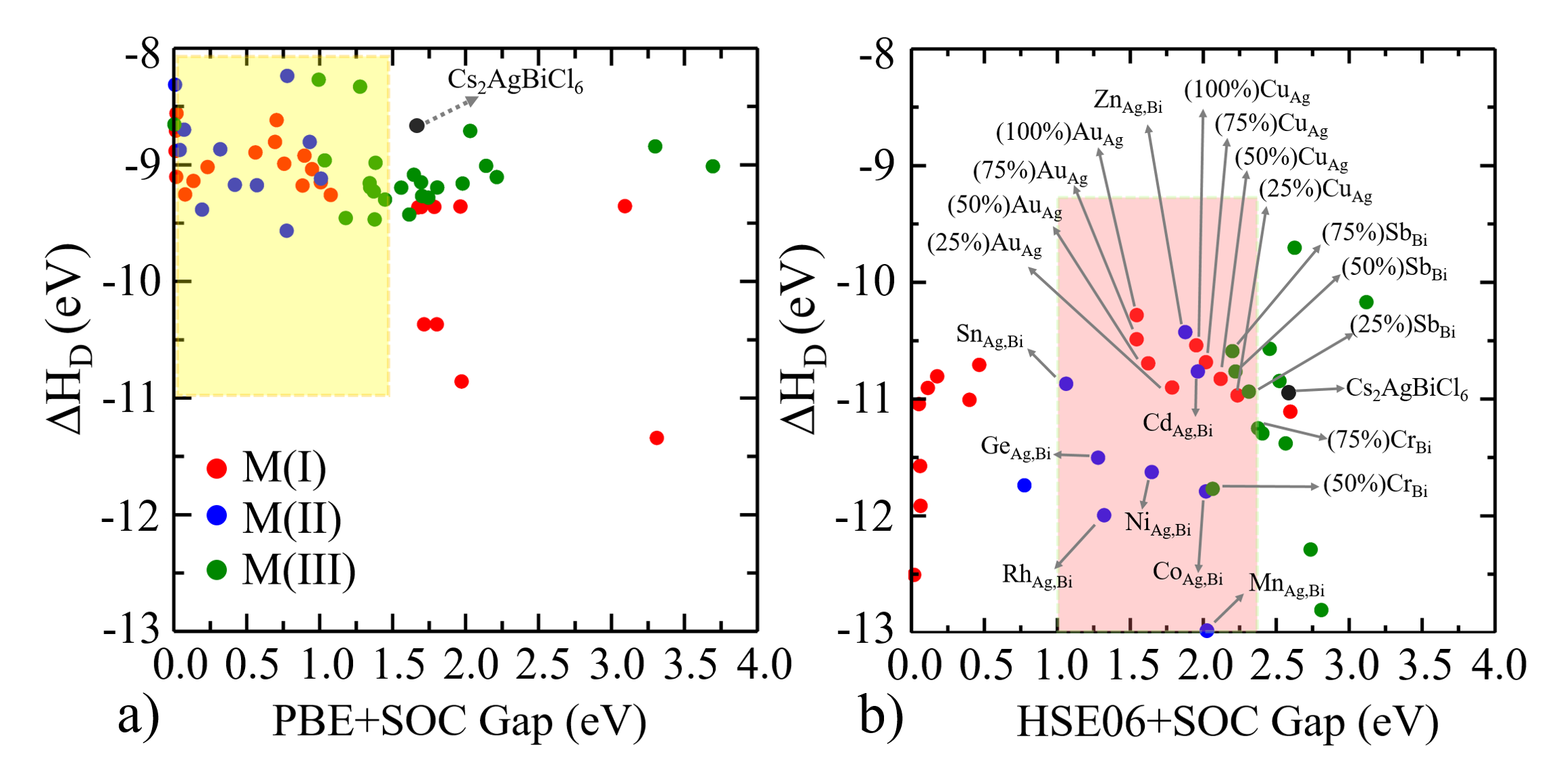}
	\caption{Band gap and $\Delta$H$_{\textrm{D}}$ using (a) PBE+SOC and (b) HSE06+SOC $\epsilon_{xc}$ functionals. (Here, red, blue and green color circular dots correspond to M(I) (e.g., substitution of $25\%$ Au at Ag-site (($25\%$)Au$_{\textrm{Ag}}$)), M(II) (e.g., substitution of Sn at Ag- and Bi-site simultaneously (Sn$_{\textrm{Ag},\textrm{Bi}}$)) and M(III) (e.g., substitution of $25\%$ Sb at Bi-site (($25\%$)Sb$_{\textrm{Bi}}$)), respectively.)}
	\label{fig1}
\end{figure}
We have started with 64 configurations of double perovskites obtained on mixing the Ag-Cl and Bi-Cl sublattices with M(I) (viz. Au, Cu, In, K, Na, and Ti), M(II) (viz. Cd, Co, Cu, Ge, Mn, Mo, Ni, Sn, V, Zn, and Rh), and M(III) (viz. Cr, Ga, In, Tl, Sb, and Y). Here, we have varied the concentration of the alloying atom viz. 25\%, 50\%, 75\% and 100\%. 
Two fundamental factors need to be satisfied for the stability of double perovskites to exist in high symmetry cubic structure, viz.  Goldschmidt's tolerance factor (\textit{t})~\cite{kieslich2015extended} and octahedral factor (\textit{$\mu$})~\cite{li2008formability}.
For structural stability, $t$ should lie between 0.8 and 1.0, and $\mu$ should be greater than 0.41~\cite{kangsabanik2018double}.
To calculate these two fundamental factors for various double perovskite configurations, we have employed a strategy.~\cite{S14}.
All the selected double perovskite configurations satisfy these stability criteria (see Table S1 in SI). After structural stability, we have also determined the thermodynamic stability by calculating $\Delta$H$_{\textrm{D}}$ for decomposition of different conformers (obtained after alloying with M(I), M(II), and M(III)) into binary compounds, using following equations:
\begin{equation}
	\begin{split}
	\Delta \textrm{H}_{\textrm{D}}(\textrm{M(I)})  = &\; \textrm{E}\left(\textrm{Cs}_8\textrm{Ag}_{4-x}\textrm{M(I)}_x\textrm{Bi}_4\textrm{Cl}_{24}\right) -  \left(4-x\right)\textrm{E}\left(\textrm{Ag}\textrm{Cl}\right) - 8\textrm{E}\left(\textrm{Cs}\textrm{Cl}\right)\\& - 4\textrm{E}\left(\textrm{Bi}\textrm{Cl}_3\right) - x\textrm{E}\left(\textrm{M(I)}\textrm{Cl}\right)
	\label{eq:1}
	\end{split}
\end{equation} 
\begin{equation}
	\begin{split}    
	\Delta \textrm{H}_{\textrm{D}}(\textrm{M(II)}) = &\; \textrm{E}\left(\textrm{Cs}_8\textrm{Ag}_{3}\textrm{M(II)}_{2}\textrm{Bi}_{3}\textrm{Cl}_{24}\right) -  3\textrm{E}\left(\textrm{Ag}\textrm{Cl}\right) - 8\textrm{E}\left(\textrm{Cs}\textrm{Cl}\right)\\ & - 3\textrm{E}\left(\textrm{Bi}\textrm{Cl}_3\right) - 2\textrm{E}\left(\textrm{M(II)}\textrm{Cl}_2\right) 
	\end{split}
\end{equation}
\begin{equation} 
	\begin{split} 
	\Delta \textrm{H}_{\textrm{D}}(\textrm{M(III)}) = &\;  \textrm{E}\left(\textrm{Cs}_8\textrm{Ag}_4\textrm{M(III)}_x\textrm{Bi}_{4-x}\textrm{Cl}_{24}\right) - 4\textrm{E}\left(\textrm{Ag}\textrm{Cl}\right) - 8\textrm{E}\left(\textrm{Cs}\textrm{Cl}\right)\\ & - \left(4-x\right)\textrm{E}\left(\textrm{Bi}\textrm{Cl}_3\right) - x\textrm{E}\left(\textrm{M(III)}\textrm{Cl}_3\right)
	\label{eq:3}
	\end{split}
\end{equation}
In Equation \ref{eq:1} and \ref{eq:3}, \textit{x} can have value 1, 2, 3 or 4.
Based on the band gap and $\Delta$H$_{\textrm{D}}$ (calculated using PBE+SOC), a pre-screening process has been employed to find the suitable configurations. In Figure \ref{fig1}(a), the promising configurations lie within the shaded region for which band gap is varying from 0.0 to 1.5 eV (for further details see Table S2, S3 and S4 in SI). 
For all the prescreened configurations of double perovskite, $\Delta$H$_{\textrm{D}}$ is negative, indicating that all the considered systems are stable and these will not decompose into respective binary components.  
Subsequently, we have calculated $\Delta$H$_{\textrm{D}}$ and band gap using HSE06+SOC $\epsilon_{xc}$ functional for aforementioned selected configurations (see Figure \ref{fig1}(b)). In Figure \ref{fig1}(b), restoring Shockley-Queisser (SQ) criterion~\cite{shockley1961detailed}, we have identified 19 double perovskite configurations for which band gap lies within 1.0 to 2.3 eV, which is relevant for solar cell and other optoelectronic devices.

Note that on increasing the concentration of the external element, if the band gap is increased (or decreased), it increases (or decreases) consistently on further increasing the concentration~\cite{MK3}. However, in some cases, we have noticed an irregular change in band gap on varying the concentration of the alloying atoms. For instance, on increasing the percentage of Sb at Bi-sites, band gap decreases up to 75\% substitution, and thereafter, an increment in band gap has been observed on 100\% substitution. Similar kind of change in band gap has also been observed on complete substitution of other elements at Ag or Bi site (see Table S2, S3 and S4 in SI for details). To understand this change in the band gap on $100\%$ substitution of Sb, we have plotted the band structures of Sb-alloyed system with different concentrations of Sb (see Figure \ref{fig2}).
\begin{figure}
	\includegraphics[width=0.8\columnwidth,clip]{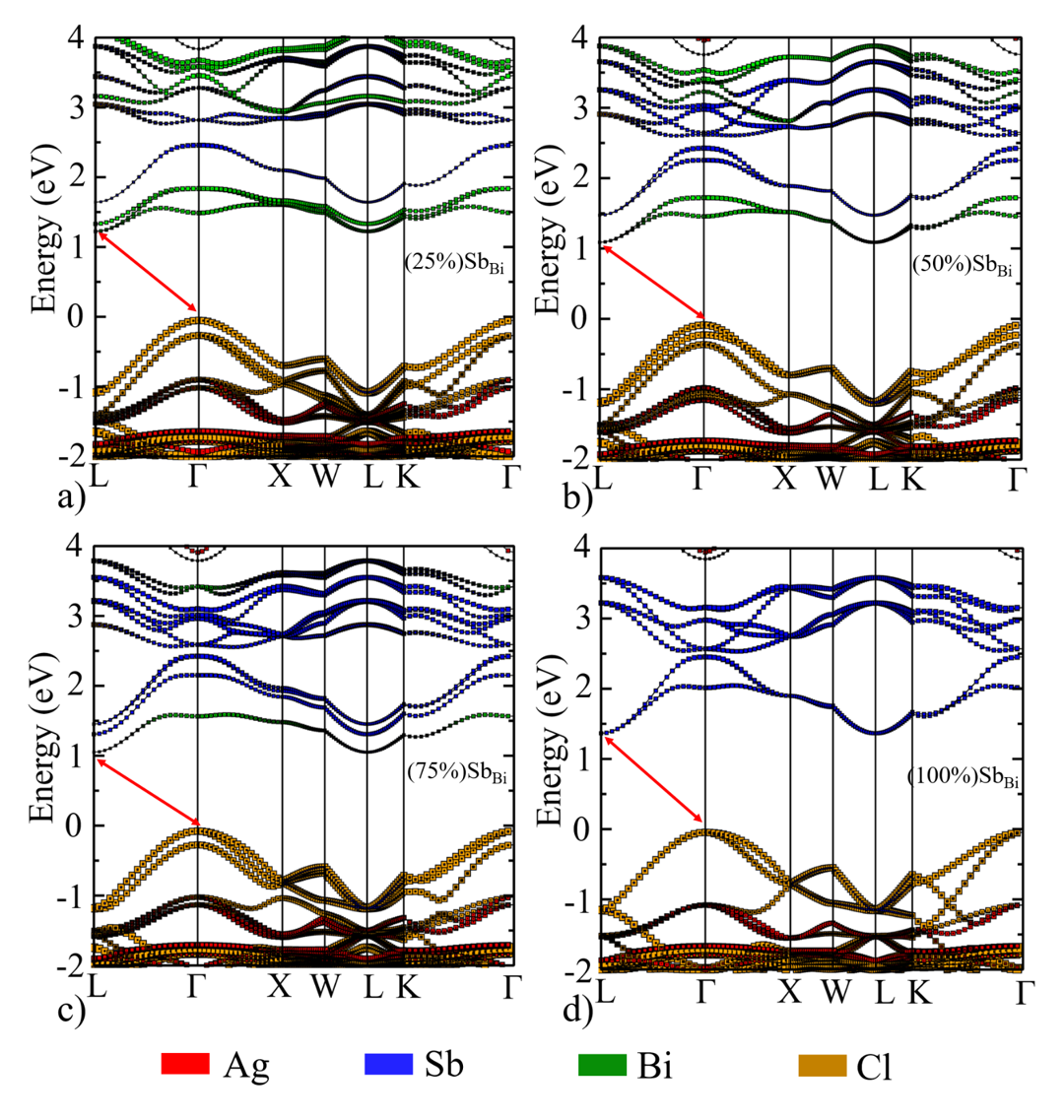}
	\caption{Band structure of (a) Cs$_8$Ag$_4$Sb$_1$Bi$_3$Cl$_{24}$, (b) Cs$_8$Ag$_4$Sb$_2$Bi$_2$Cl$_{24}$, (c) Cs$_8$Ag$_4$Sb$_3$Bi$_1$Cl$_{24}$ and (d) Cs$_8$Ag$_4$Sb$_4$Cl$_{24}$ using PBE+SOC $\epsilon_{xc}$ functional.}
	\label{fig2}
\end{figure}
We can clearly see that on $100\%$ substitution of Sb at Bi-sites, the lowest energy level in the conduction band corresponding to Bi-orbitals (that is present in Figure \ref{fig2}(a-c)) disappears. Consequently, there is an increment in the band gap (see Figure \ref{fig2}(d)). In addition, from Figure \ref{fig2}, we can see that SOC effect is attributed to the fundamental mismatch of Ag- and Bi-orbitals, which is also mentioned by Savory et al.~\cite{savory2016can}. Similar SOC effect can be observed from band structures for Au substitution at Ag-sites as well (see Figure S3 in SI). Hence, in some cases, on complete substitution either at Ag- or Bi-sites, there is an inconsistent change in the band gap (i.e., on complete substitution of Ag with Na, K etc., similar inconsistency has been observed). For more details also see Figure S4 and S5.    

\begin{figure}
	\includegraphics[width=1.0\columnwidth,clip]{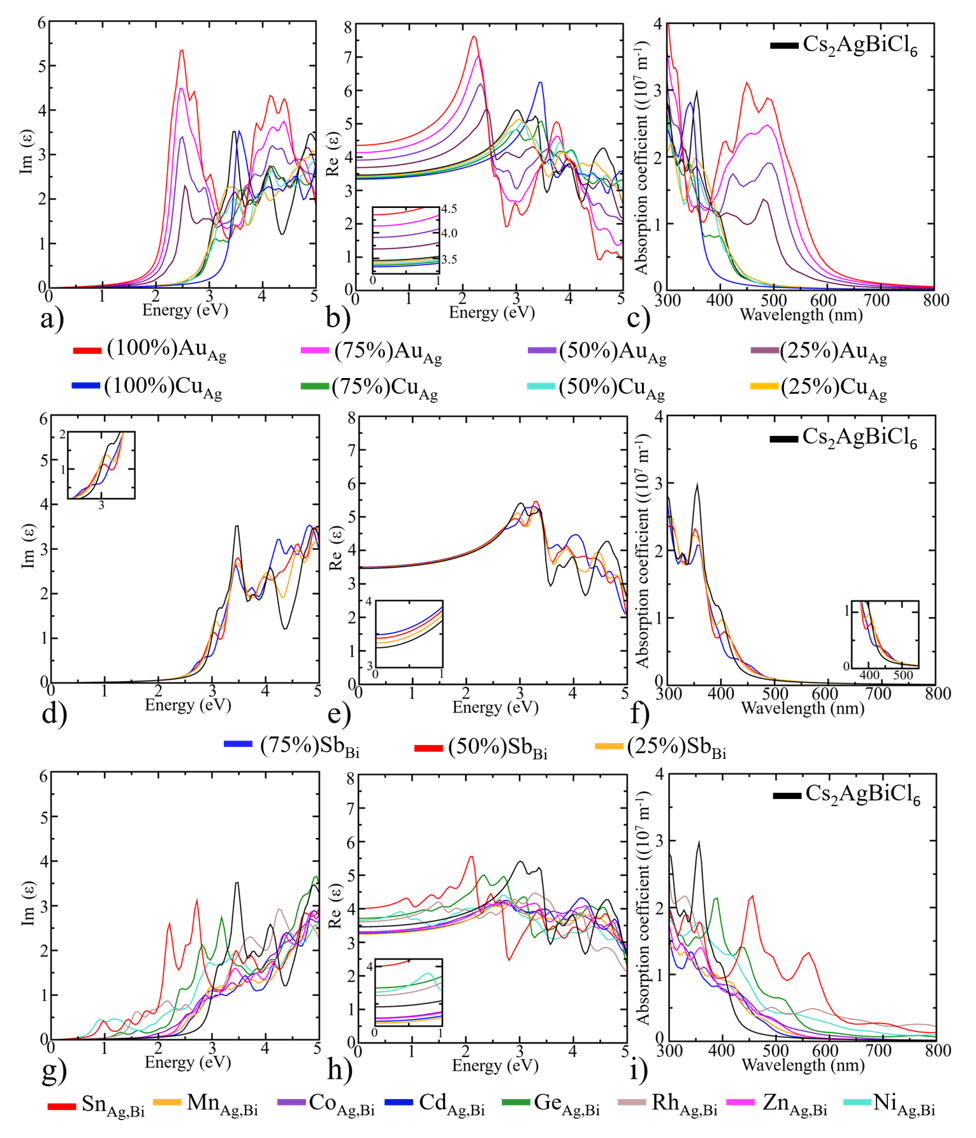}
	\caption{Variation of imaginary part of dielectric constant (Im($\varepsilon$)) of Cs$_2$AgBiCl$_6$ sublattice mixed with (a) monovalent (M(I)), (d) divalent (M(II)) and (g) trivalent (M(III)) cations, respectively. Variation of real part of dielectric constant (Re($\varepsilon$)) of Cs$_2$AgBiCl$_6$ sublattice mixed with (b) monovalent (M(I)), (e) divalent (M(II)) and (h) trivalent (M(III)) cations, respectively. Absorption coefficient of Cs$_2$AgBiCl$_6$ sublattice mixed with (c)  monovalent (M(I)), (f) divalent (M(II)) and (i) trivalent (M(III)) cations, respectively. Note that all calculations have been done using HSE06+SOC $\epsilon_{xc}$ functional.}
	\label{fig3} 
\end{figure}

Next, we have determined the optical properties of the 19 potential candidate structures (viz. Cs$_8$Ag$_{4-x}$M(I)$_x$Bi$_4$Cl$_{24}$ (M(I) = Au and Cu); $x\in[1,4]$), Cs$_8$Ag$_3$M(II)$_2$Bi$_3$Cl$_{24}$ (M(II = Sn, Ge, Rh, Ni, Co, Cd, Mn and Zn), and Cs$_8$Ag$_{4}$Sb$_x$Bi$_{4-x}$Cl$_{24}$ ($x\epsilon[1,3]$), which possess band gap in an appropriate range for solar cell and other optoelectronic applications.
To determine optical properties, frequency dependent complex dielectric function, $\varepsilon(\omega)$ = Re($\varepsilon$) + Im($\varepsilon$) has been calculated using HSE06+SOC $\epsilon_{xc}$ functional as shown in Figure \ref{fig3} (the results of optical properties without SOC are shown in Figure S6 and S7 in SI for bechmark purpose).
Therefore, using Re($\varepsilon$) and Im($\varepsilon$) of dielectric function, various optical properties, e.g., refractive index ($\eta$), extinction coefficient ($\kappa$), and absorption coefficient ($\alpha$) can be computed using following expressions:
\begin{align}
\eta&= \frac{1}{\sqrt{2}}\left[\sqrt{\textrm{Re}(\varepsilon)^2 + \textrm{Im}(\varepsilon)^2} + \textrm{Re}(\varepsilon)\right]^{\frac{1}{2}}
\label{eq:4}\\
\kappa&= \frac{1}{\sqrt{2}}\left[\sqrt{\textrm{Re}(\varepsilon)^2 + \textrm{Im}(\varepsilon)^2} - \textrm{Re}(\varepsilon)\right]^{\frac{1}{2}}
\label{eq:5}\\
\alpha &= \frac{2\omega\kappa}{c} 
\label{eq:6}
\end{align}
Here, in Equation \ref{eq:6}, $\omega$ and c correspond to angular frequency and speed of light, respectively.
In Figure \ref{fig3}(a), peaks of conformers are red shifted w.r.t. Cs$_2$AgBiCl$_6$. Here, red shift is attributed to the reduction of the band gap. There is more red shift on increasing the concentration of Au in comparison to Cu. On the other hand, the real part of the dielectric constant ($\omega=0$) increases, with increase in the concentration of Au and decreases, with increase in the concentration of Cu (see Figure \ref{fig3}(b)). For high degree of charge screening, which can prohibit radiative electron-hole recombination, a large value of Re($\omega$) at $\omega=0$ is indispensable~\cite{basera2019self}. Hence, the solar cell absorber, which exhibits large Re($\omega$) at $\omega=0$ is more efficient. In view of this, Au substitution is more beneficial than Cu for replacing Ag-sites.

In Figure \ref{fig3}(c), we can clearly see that peaks corresponding to Au substitution are red shifted w.r.t. Cs$_2$AgBiCl$_6$ and show good optical absorption within visible region. Hence, we have discerned that in case of alloying with M(I), substitution of Au at Ag-sites acts as a promising candidate rather than Cu. Likewise, in case of alloying with M(III), substitution with Sb at Bi-sites acts as a promising candidate. From Figure \ref{fig3}(d), the red shift w.r.t. pristine (see inset to have a clear view) conveys that band gap decreases on increasing the concentration of Sb. However, for $100\%$ Sb$_{\textrm{Bi}}$, band gap increases, which can be seen from band structure (see Figure \ref{fig2}(d)). From Figure \ref{fig3}(e) and \ref{fig3}(f), we can see that optical properties are enhanced on increasing the concentration of Sb (upto 75\%) w.r.t. pristine. Similarly, in case of alloying with M(II), from Figure \ref{fig3}(g) and \ref{fig3}(i), there is a red shift of absorption peak w.r.t. Cs$_2$AgBiCl$_6$. Moreover, static value of Re($\omega$) at $\omega=0$ for Sn$_{\textrm{Ag},\textrm{Bi}}$, Rh$_{\textrm{Ag},\textrm{Bi}}$, Ni$_{\textrm{Ag},\textrm{Bi}}$ and Ge$_{\textrm{Ag},\textrm{Bi}}$ is larger than pristine system; but for other alloyed systems (viz. Zn$_{\textrm{Ag},\textrm{Bi}}$, Co$_{\textrm{Ag},\textrm{Bi}}$, and Mn$_{\textrm{Ag},\textrm{Bi}}$), it is lower than pristine system (see Figure \ref{fig3}(h)). Out of all the M(II) selected candidates, these four (viz. Sn, Rh, Ni and Ge) are the best aspirants. Although, only Sn$_{\textrm{Ag},\textrm{Bi}}$ and Ge$_{\textrm{Ag},\textrm{Bi}}$  show direct band gap, the other two are to be more suitable for optoelectronic devices excluding solar cell. On the other hand, since Sn is cheaper than Ge, it serves as better candidate for alloying. It has been reported that complete substitution of Sn degrades the properties of the system~\cite{S1,S2}. Therefore, in order to overcome this problem, we have done partial substitution of Sn. Also, from Figure~\ref{fig3}, we can compare the scales (viz. values of dielectric constants and peaks of absorption coefficient in visible region) and infer that alloying with M(II) is a better choice to enhance the optical properties.
\begin{figure}
	\includegraphics[width=0.5\columnwidth,clip]{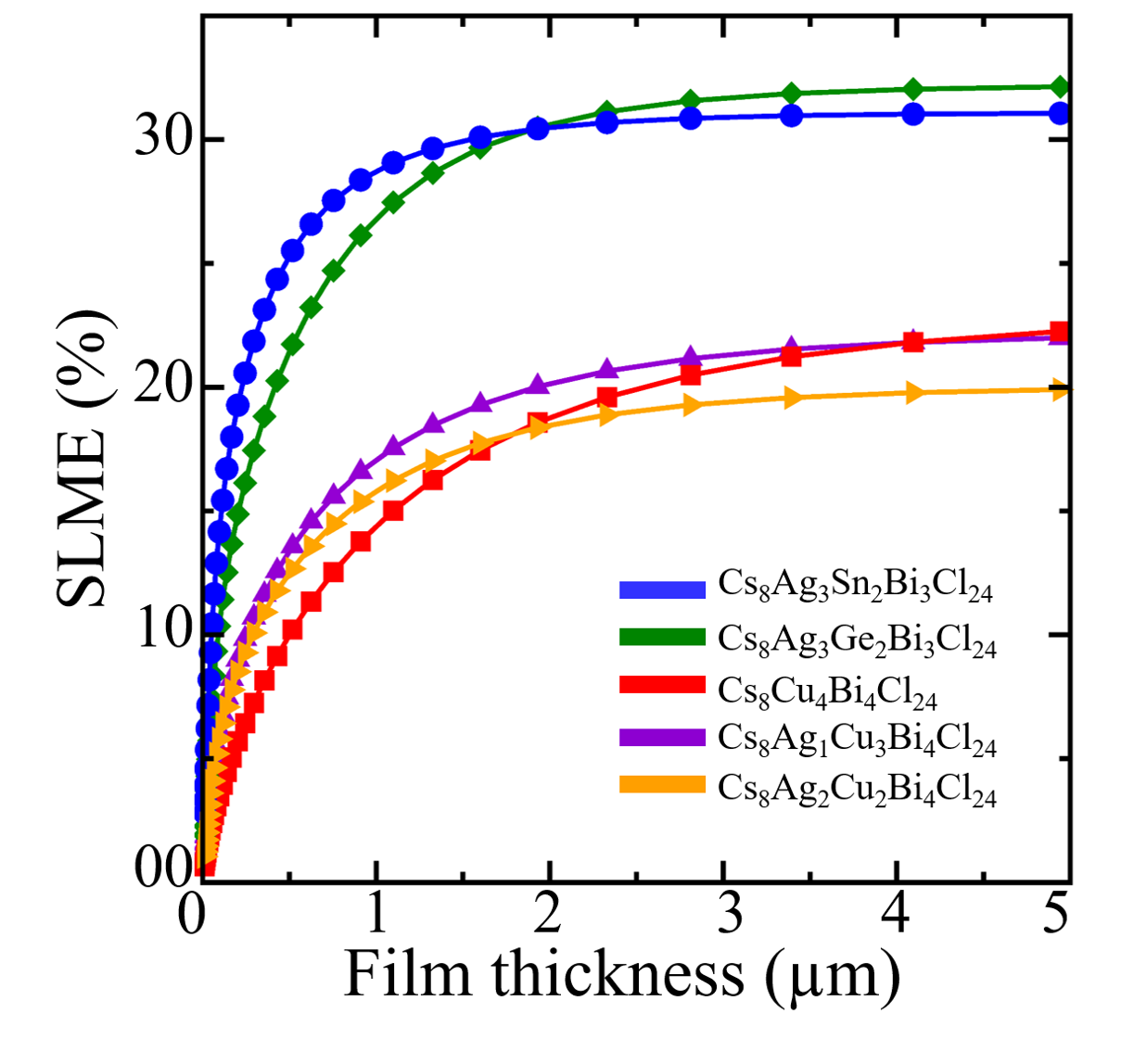}
	\caption{Variation of SLME w.r.t. the thickness of solar cell absorber.}
	\label{fig4}
\end{figure}
On comparing all the above results in Figure \ref{fig1} and \ref{fig3}, we have revealed that partial Sn substitution acts as a promising candidate to enhance the optical properties of Cs$_2$AgBiCl$_6$ without degrading the stability. In addition, it also exhibits direct band gap property (see Figure S4 in SI). Thus Sn$_{\textrm{Ag},\textrm{Bi}}$ acts as a rational candidate for solar cell application.

Lastly, to design highly efficient solar cell absorber, spectroscopic limited maximum efficiency (SLME)~\cite{kangsabanik2018double} has been calculated. The SLME is based on the improved Shockley-Queisser model. It depends on the absorption coefficient, thickness of the film absorber and nature of the band gap of the material. For an efficient solar cell, Yu et al.~\cite{yu2012identification} have proposed the concept of SLME based on Fermi Golden rule. According to Fermi Golden rule, the optical absorption is directly proportional to
\begin{align}
    \frac{2\pi}{\hbar}\int|\left<\nu|\hat{H}|c\right>|^2\frac{2}{8\pi^3}\delta\left(E_c(\overrightarrow{k}) - E_\nu(\overrightarrow{k}) - \hbar\omega\right)d^3k
   \label{eq:7}
\end{align} 
where $\left<\nu|\hat{H}|c\right>$ is the transition matrix. $\hbar\omega$ is the photonic energy required for transition from states in valence band ($\nu$) to the states in conduction band (c). In Equation \ref{eq:7}, integration goes over the whole reciprocal space. 
We have calculated the SLME of those alloyed systems, which possess direct band gap, as a function of thickness of the absorber layer (see Figure \ref{fig4}). Here, in Table \ref{Table1}, we have shown SLME of few double perovskites at 5 $\mu$m and compared our results with other efficient hybrid perovskites that are reported recently.
\begin{table}
	\caption{Comparison of SLME of double perovskites with hybrid perovskite at 5 $\mu$m absorber layer thickness.}
	\begin{center}
		\begin{tabular}{|c|c|} \hline
			Conformers & SLME (\%) \\ \hline
			Cs$_8$Ag$_{3}$Ge$_{2}$Bi$_{3}$Cl$_{24}$ & 32.08  \\ \hline
			Cs$_8$Ag$_{3}$Sn$_{2}$Bi$_{3}$Cl$_{24}$  & 30.91  \\ \hline
			Cs$_8$Cu$_4$Bi$_4$Cl$_{24}$ & 22.24 \\ \hline
			Cs$_8$Ag$_{1}$Cu$_{3}$Bi$_4$Cl$_{24}$ & 21.85 \\ \hline  
			Cs$_8$Ag$_{2}$Cu$_{2}$Bi$_4$Cl$_{24}$ & 19.80 \\ \hline
			MA$_8$Pb$_8$I$_{24}$  &  31.02~\cite{PhysRevB.101.054108}  \\ \hline
			MA$_8$Pb$_7$Sn$_1$I$_{24}$   &  33.02~\cite{PhysRevB.101.054108}  \\ \hline
			FA$_8$Pb$_4$Sn$_4$Br$_{24}$ & 26.74~\cite{kk}\\ \hline 
		\end{tabular}
		\label{Table1}
	\end{center}
	\label{table:1}
\end{table}
From Table \ref{Table1}, SLME of Cs$_8$Ag$_{3}$Ge$_{2}$Bi$_{3}$Cl$_{24}$ and Cs$_8$Ag$_{3}$Sn$_{2}$Bi$_{3}$Cl$_{24}$ are 32.08\% and 30.91\%, respectively. These numbers are very much encouraging from application perspective in solar cells. In addition, they are more stable, while in contact with air and moisture as compared to IO hybrid perovskites~\cite{mcclure2016cs2agbix6,filip2016band}. 

In summary, we have presented a thorough study of alloying double perovskite Cs$_2$AgBiCl$_6$  with M(I), M(II) and M(III) cations. 
The role of SOC is important to accurately predict the band gap and band-edge positions in such systems. All the mixed sublattices are structurally stable as indicated by the Goldschmidt's tolerance factor and octahedral factor. The enthalpies of decomposition are negative, indicating the thermodynamic stability of alloyed systems. 
We have revealed that for substitution at Bi-sites, on increasing the concentration of Sb, band gap decreases upto 75\% Sb$_{\textrm{Bi}}$ substitution. However, there is a sudden increase in band gap for 100\% Sb$_{\textrm{Bi}}$ due to complete removal of Bi. We have also identified that Au substitution can enhance the optical properties effectively due to reduction in band gap. Hence, we have concluded that partial substitution of Au and Sb at Ag- and Bi-sites, respectively, will be cost-effective and efficient to enhance the optical properties. Out of alloying with M(I), M(II) and M(III), M(II) (Sn$_{\textrm{Ag},\textrm{Bi}}$, Rh$_{\textrm{Ag},\textrm{Bi}}$, Ni$_{\textrm{Ag},\textrm{Bi}}$ and Ge$_{\textrm{Ag},\textrm{Bi}}$) substitutions come out to be superior for optical properties. However, only in case of Sn$_{\textrm{Ag},\textrm{Bi}}$ and Ge$_{\textrm{Ag},\textrm{Bi}}$, direct band gaps are noticed. SLME of Cs$_8$Ag$_{3}$Ge$_{2}$Bi$_{3}$Cl$_{24}$ and Cs$_8$Ag$_{3}$Sn$_{2}$Bi$_{3}$Cl$_{24}$ are 32.08\% and 30.91\%, respectively, which definitely suggest huge promise towards prospective solar cell absorbers. 


\section{Computational Methods}
First-principles calculations have been performed using DFT with PAW pseudopotential method~\cite{hohenberg1964inhomogeneous, kohn1965self, blochl1994projector} as implemented in Vienna \textit{ab initio} simulation package (VASP)~\cite{kresse199614251}. Cs$_{2}$AgBiCl$_{6}$ is a cubic structure having space group \textit{Pnma}. It comprises of 40 atoms (4 formula units) in the unit cell i.e., Cs$_8$Ag$_4$Bi$_4$Cl$_{24}$ and single defect state remains fully localized with periodic boundary conditions. We have used exchange-correlation ($\epsilon_{xc}$) functionals viz. GGA (PBE~\cite{perdew1992atoms}) and hybrid functional HSE06~\cite{heyd2003hybrid} with and without SOC for the calculations. The total energy tolerance is set to 0.001 meV. The Hellmann-Feynman forces~\cite{pulay1980convergence} have been converged upto 0.001 eV/$\textrm{\AA}$ by conjugate gradient (CG) minimization to obtain optimized ground state structures. The k-mesh has been generated by the Monkhorst-Pack~\cite{monkhorst1976special} method. All the structures are fully relaxed with k-mesh 2$\times$2$\times$2. For single-point energy calculation, k-mesh is converged and kept fixed at 5$\times$5$\times$5. The plane wave energy cut-off is set to 600 eV in our calculations. 
\begin{acknowledgement}
DG acknowledges UGC, India, for the junior research fellowship [grant no.: 1268/(CSIR-UGC NET JUNE 2018)]. PB acknowledges UGC, India, for the junior research fellowship [grant no.: 1392/(CSIR-UGC NET JUNE 2018)]. MK acknowledges CSIR, India, for the senior research fellowship [grant no.: 09/086(1292)/2017-EMR-I]. SB acknowledges the financial support from SERB under core research grant (grant no. CRG/2019/000647). We acknowledge the High Performance Computing (HPC) facility at IIT Delhi for computational resources.
\end{acknowledgement}
\begin{suppinfo}
Details regarding band gap using different functionals for various conformers, optical properties using HSE06, and partial density of states (pDOS) have been given in the supporting information file. 
\end{suppinfo}

\begin{center}
{\Large \bf Supplemental Material}\\ 
\end{center}
\begin{enumerate}[\bf I.]
\item Band gap, tolerance factor, octahedral factor and enthalpy of decomposition of different configurations 
\item Band structures for Au and Sn substitution
\item Partial density of states (pDOS) plot, showing contribution of various orbitals in valence band maximum (VBM) and conduction band minimum (CBm) of few selected conformers
\item Optical properties using HSE06 $\epsilon_{xc}$ functional \end{enumerate}
\vspace*{12pt}
\newpage
\section{{Band gap, tolerance factor, octahedral factor and enthalpy of decomposition of different configurations}}

\begin{figure}[H]
    \centering
	\includegraphics[width=1.0\columnwidth,clip]{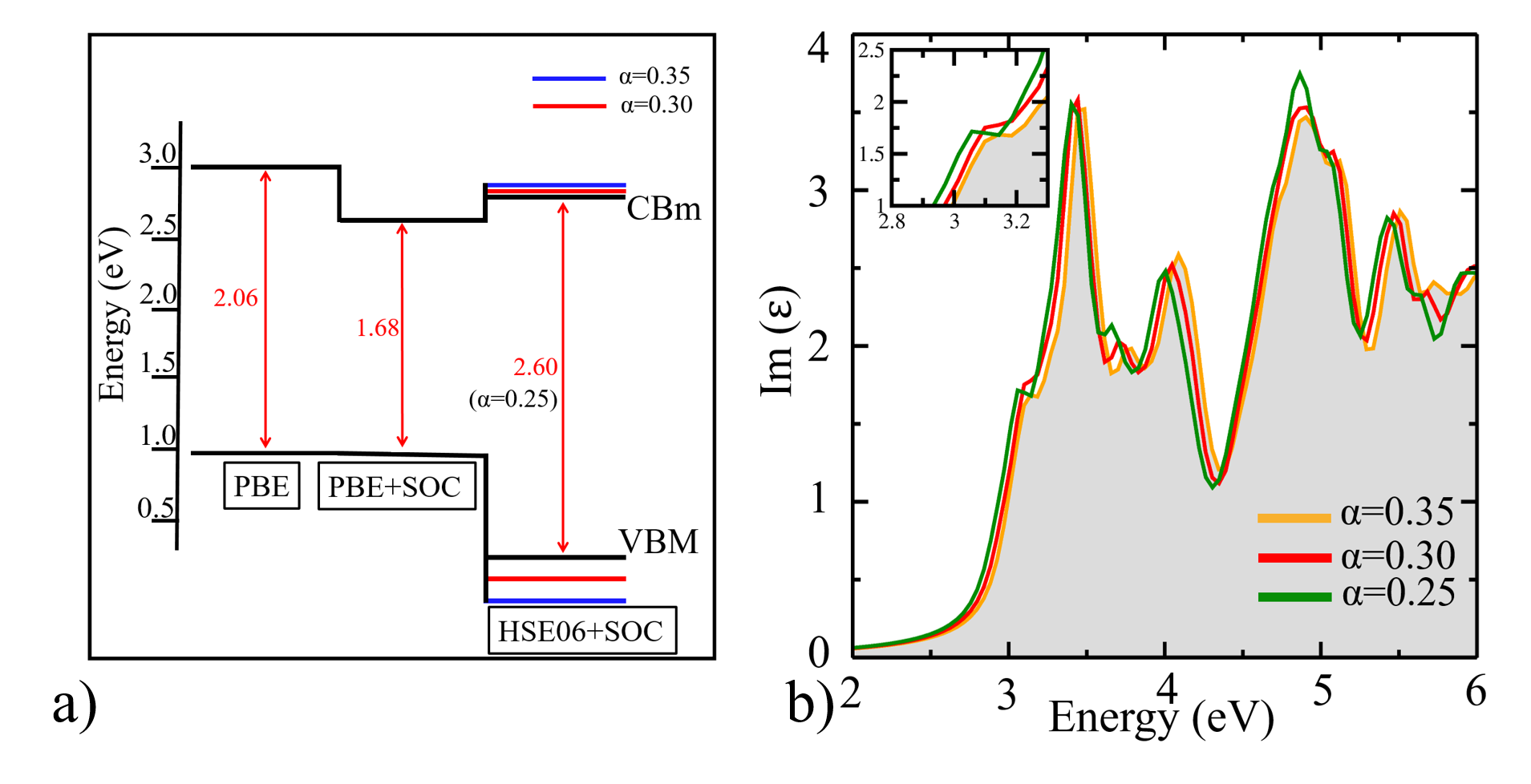}
	\caption{(a) Band edge alignment of VBM and CBm with PBE, PBE+SOC and HSE06+SOC and (b) Absorption spectra of Cs$_2$AgBiCl$_6$ using HSE06+SOC $\epsilon_{xc}$ functional, where band gaps of 2.60, 2.79 and 2.99 eV have been obtained with Hartree-Fock exchange fraction ($\alpha$) = 0.25, 0.30 and 0.35, respectively.}
	\label{fig2}
\end{figure}

\begin{figure}[H]
	\centering
	\includegraphics[width=0.8\columnwidth,clip]{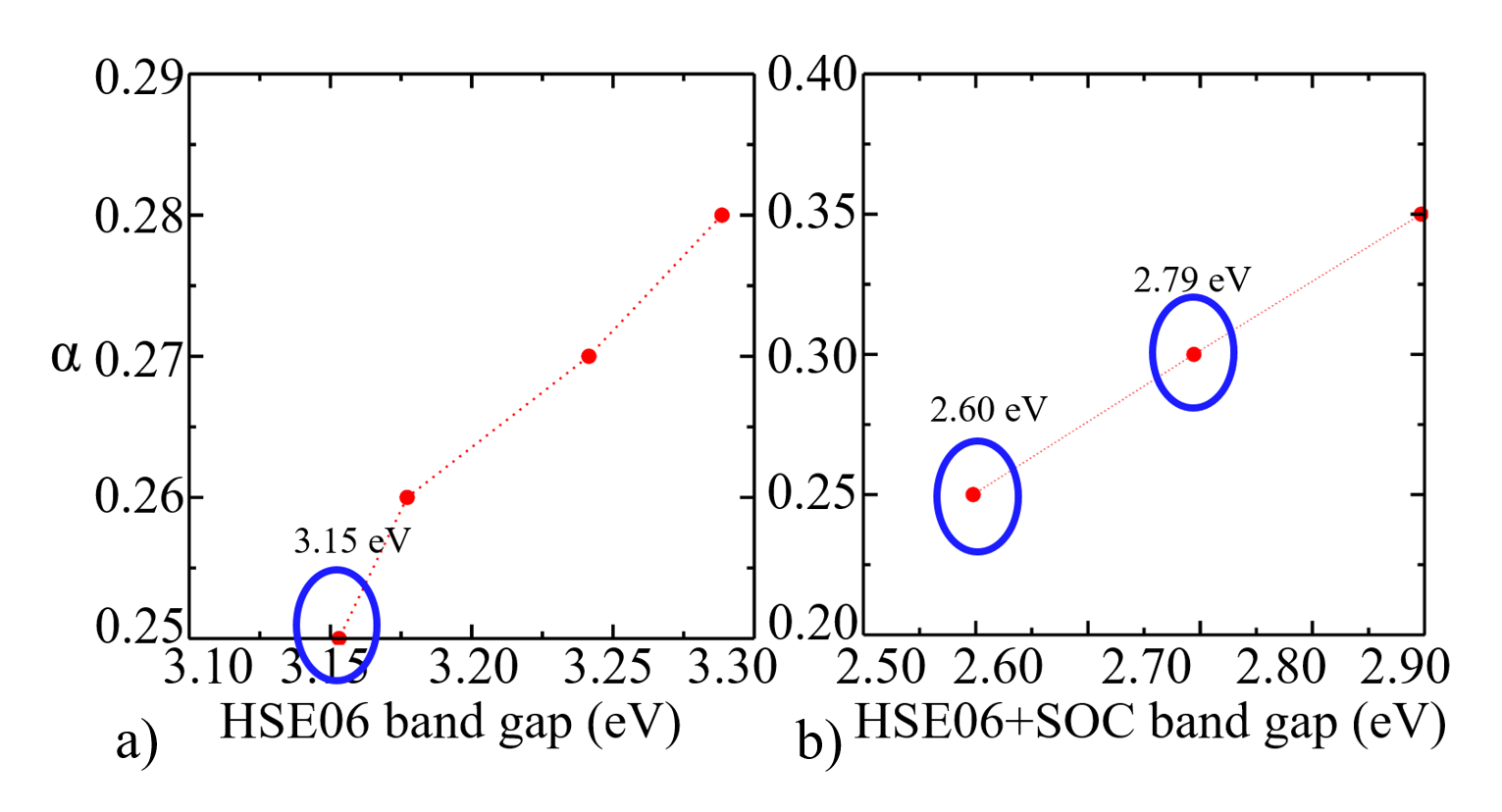}
	\caption{Band gap of Cs$_2$AgBiCl$_6$, using HSE06 and HSE06+SOC $\epsilon_{xc}$ functionals with different values of Hartree-Fock exchange fraction ($\alpha$).}
	\label{fig1}
\end{figure}
In Figure \ref{fig2}(a), shifting of band edges viz. conduction band minimum (CBm) and valence band maximum (VBM) positions on implementation of different $\epsilon_{xc}$ functionals viz. PBE, PBE+SOC and HSE06+SOC have been shown. Correct positions of band edges are mandatory for prediction of excited state properties of a material. An estimated band gap of Cs$_2$AgBiCl$_6$ with PBE is 2.06 eV, which is not in agreement with experimental value of 2.77 eV. Next, spin-orbit coupling (SOC) effect has been incorporated. It lowers the CBm owing to splitting of degenerate levels, without altering the VBM. Hence, PBE+SOC gives reduced band gap value (see Figure \ref{fig2}(a)). Apart from this, the presence of self-interaction error in PBE leads to an incorrect position of VBM. Partial elimination of self-interaction error can be achieved on incorporating hybrid $\epsilon_{xc}$ functional that takes into account Hartree-Fock exchange fraction ($\alpha$). With different values of $\alpha$ i.e. $0.25$, $0.30$ and $0.35$, the obtained band gaps are $2.60$ eV, $2.79$ eV and $2.99$ eV, respectively (see Figure \ref{fig2}(b)). The band gap diverges from the experimental value on increasing $\alpha$ beyond $0.30$. The default value of $\alpha$ i.e., $0.25$ also gives band gap in good approximation with the experimental value. In Figure \ref{fig2}(b), the imaginary part of the dielectric function corresponding to three different values of $\alpha$ = $0.25$, $0.30$ and $0.35$ has been calculated.  From the positions of three peaks, it is confirmed that on increasing the value of $\alpha$, band gap increases.
\begin{longtable}{p{4.2cm} p{2.0cm}p{2.0cm}} 
		\caption{Tolerance and octahedral factor of different conformers}\\ \hline
		Conformers & Tolerance factor ($t$) & Octahedral factor ($\mu$) \\ \hline 
		Cs$_8$Ag$_4$Bi$_4$Cl$_{24}$& 0.90 &0.60 \\ \hline
		Cs$_8$Ag$_3$Au$_1$Bi$_4$Cl$_{24}$&0.89 &0.62\\ \hline
		Cs$_8$Ag$_2$Au$_2$Bi$_4$Cl$_{24}$& 0.88 &0.63\\ \hline
		Cs$_8$Ag$_1$Au$_3$Bi$_4$Cl$_{24}$&0.87& 0.65\\ \hline
		Cs$_8$Au$_4$Bi$_4$Cl$_{24}$&0.87 &0.66\\ \hline
		Cs$_8$Ag$_3$Cu$_1$Bi$_4$Cl$_{24}$&0.91& 0.58\\ \hline
		Cs$_8$Ag$_2$Cu$_2$Bi$_4$Cl$_{24}$&0.93& 0.55\\ \hline
		Cs$_8$Ag$_1$Cu$_3$Bi$_4$Cl$_{24}$&0.95 &0.52\\ \hline
		Cs$_8$Cu$_4$Bi$_4$Cl$_{24}$&0.96 &0.50\\ \hline
		Cs$_8$Ag$_3$In$_1$Bi$_4$Cl$_{24}$&0.91 & 0.58\\ \hline
		Cs$_8$Ag$_2$In$_2$Bi$_4$Cl$_{24}$&0.93 &0.55\\ \hline
		Cs$_8$Ag$_1$In$_3$Bi$_4$Cl$_{24}$&0.94&0.53\\ \hline
		Cs$_8$In$_4$Bi$_4$Cl$_{24}$&0.96 &0.51\\ \hline
		Cs$_8$Ag$_3$K$_1$Bi$_4$Cl$_{24}$&0.89 &0.62\\ \hline
		Cs$_8$Ag$_2$K$_2$Bi$_4$Cl$_{24}$&0.88 &0.63\\ \hline
		Cs$_8$Ag$_1$K$_3$Bi$_4$Cl$_{24}$&0.87 &0.65\\ \hline
		Cs$_8$K$_4$Bi$_4$Cl$_{24}$&0.87 &0.66\\ \hline
		Cs$_8$Ag$_3$Na$_1$Bi$_4$Cl$_{24}$&0.90 &0.59\\ \hline
		Cs$_8$Ag$_2$Na$_2$Bi$_4$Cl$_{24}$&0.91 &0.58\\ \hline
		Cs$_8$Ag$_1$Na$_3$Bi$_4$Cl$_{24}$&0.92 &0.58\\ \hline
		Cs$_8$Na$_4$Bi$_4$Cl$_{24}$&0.92 &0.57\\ \hline
		Cs$_8$Ag$_3$Ti$_1$Bi$_4$Cl$_{24}$&0.91 &0.58\\ \hline
		Cs$_8$Ag$_2$Ti$_2$Bi$_4$Cl$_{24}$&0.92 &0.56\\ \hline
		Cs$_8$Ag$_1$Ti$_3$Bi$_4$Cl$_{24}$&0.93  &0.54\\ \hline
		Cs$_8$Ti$_4$Bi$_4$Cl$_{24}$&0.95 &0.52\\ \hline
		Cs$_8$Ag$_4$Cr$_1$Bi$_3$Cl$_{24}$&0.91 &0.59\\ \hline
		Cs$_8$Ag$_4$Cr$_2$Bi$_2$Cl$_{24}$&0.92 &0.57\\ \hline
		Cs$_8$Ag$_4$Cr$_3$Bi$_1$Cl$_{24}$&0.93 &0.55\\ \hline
		Cs$_8$Ag$_4$Cr$_4$Cl$_{24}$&0.94 &0.54\\ \hline
		Cs$_8$Ag$_4$Ga$_1$Bi$_3$Cl$_{24}$&0.92 &0.57\\ \hline
		Cs$_8$Ag$_4$Ga$_2$Bi$_2$Cl$_{24}$&0.93 &0.55\\ \hline
		Cs$_8$Ag$_4$Ga$_3$Bi$_1$Cl$_{24}$&0.95 &0.52\\ \hline
		Cs$_8$Ag$_4$Ga$_4$Cl$_{24}$&0.97 &0.49\\ \hline
		Cs$_8$Ag$_4$In$_1$Bi$_3$Cl$_{24}$&0.91 &0.59\\ \hline
		Cs$_8$Ag$_4$In$_2$Bi$_2$Cl$_{24}$&0.92 &0.57\\ \hline
		Cs$_8$Ag$_4$In$_3$Bi$_1$Cl$_{24}$&0.93 &0.55\\ \hline
		Cs$_8$Ag$_4$In$_4$Cl$_{24}$&0.94 & 0.54\\ \hline
		Cs$_8$Ag$_4$Sb$_1$Bi$_3$Cl$_{24}$&0.91 & 0.58\\ \hline
		Cs$_8$Ag$_4$Sb$_2$Bi$_2$Cl$_{24}$&0.92 &0.56\\ \hline
		Cs$_8$Ag$_4$Sb$_3$Bi$_1$Cl$_{24}$&0.93  &0.55\\ \hline
		Cs$_8$Ag$_4$Sb$_4$Cl$_{24}$&0.94 &0.53\\ \hline
		Cs$_8$Ag$_4$Sc$_1$Bi$_3$Cl$_{24}$&0.91 &0.58\\ \hline
		Cs$_8$Ag$_4$Sc$_2$Bi$_2$Cl$_{24}$&0.92 &0.56\\ \hline
		Cs$_8$Ag$_4$Sc$_3$Bi$_1$Cl$_{24}$&0.93 &0.54\\ \hline
		Cs$_8$Ag$_4$Sc$_4$Cl$_{24}$&0.95 &0.52\\ \hline
		Cs$_8$Ag$_4$Tl$_1$Bi$_3$Cl$_{24}$&0.91 &0.60\\ \hline
		Cs$_8$Ag$_4$Tl$_2$Bi$_2$Cl$_{24}$&0.91 &0.58\\ \hline
		Cs$_8$Ag$_4$Tl$_3$Bi$_1$Cl$_{24}$&0.92 &0.57\\ \hline
		Cs$_8$Ag$_4$Tl$_4$Cl$_{24}$&0.92 &0.56\\ \hline
		Cs$_8$Ag$_4$Y$_1$Bi$_3$Cl$_{24}$&0.90 &0.59\\ \hline
		Cs$_8$Ag$_4$Y$_2$Bi$_2$Cl$_{24}$&0.91 &0.58\\ \hline
		Cs$_8$Ag$_4$Y$_3$Bi$_1$Cl$_{24}$&0.92 &0.58\\ \hline
		Cs$_8$Ag$_4$Y$_4$Cl$_{24}$&0.92 &0.57\\ \hline
		Cs$_8$Ag$_3$Cd$_2$Bi$_3$Cl$_{24}$&0.91 &0.58\\ \hline
		Cs$_8$Ag$_3$Co$_2$Bi$_3$Cl$_{24}$&0.93 &0.55\\ \hline
		Cs$_8$Ag$_3$Cu$_2$Bi$_3$Cl$_{24}$&0.93 &0.55\\ \hline
		Cs$_8$Ag$_3$Ge$_2$Bi$_3$Cl$_{24}$&0.93 &0.55\\ \hline
		Cs$_8$Ag$_3$Mn$_2$Bi$_3$Cl$_{24}$&0.93&0.54\\ \hline
		Cs$_8$Ag$_3$Ni$_2$Bi$_3$Cl$_{24}$&0.94 &0.54\\ \hline
		Cs$_8$Ag$_3$Sn$_2$Bi$_3$Cl$_{24}$&0.90 &0.61\\ \hline
		Cs$_8$Ag$_3$V$_2$Bi$_3$Cl$_{24}$&0.92 &0.56\\ \hline
		Cs$_8$Ag$_3$Zn$_2$Bi$_3$Cl$_{24}$&0.93 &0.55\\ \hline
		Cs$_8$Ag$_3$Rh$_2$Bi$_3$Cl$_{24}$&0.93 &0.55\\ \hline
		\label{Table4}
\end{longtable}
\newpage
\begin{table}
\caption{Band gap and enthalpy of decomposition of different conformers (double perovskites) for M(III) substitution}
\centering
\begin{tabular}{p{3.8cm} p{1.2cm}p{1.2cm}p{1.2cm}p{1.2cm}p{2.2cm}p{2.4cm}} \hline
	Conformers & PBE (eV)& PBE +SOC (eV) & HSE06 +SOC (eV)& HSE06 (eV)& $\Delta$H$_{\textrm{D}}$ (PBE+SOC) (eV) &$\Delta$H$_{\textrm{D}}$ (HSE06+SOC) (eV)\\ \hline
	Cs$_8$Ag$_4$Cr$_{1}$Bi$_{3}$Cl$_{24}$&1.47&1.34&2.37&-&-9.19&	-11.25
	\\ \hline
	Cs$_8$Ag$_4$Cr$_{2}$Bi$_{2}$Cl$_{24}$&1.10&1.03&2.06&2.43& -8.96&	-11.77
	\\ \hline
	Cs$_8$Ag$_4$Cr$_{3}$Bi$_{1}$Cl$_{24}$&0.78&0.77&2.73&2.86&-8.65 &-12.29
	\\ \hline
	Cs$_8$Ag$_4$Cr$_4$Cl$_{24}$ &0.80&0.79&2.81&2.94&-8.27&-12.81\\ \hline
	Cs$_8$Ag$_4$Ga$_{1}$Bi$_{3}$Cl$_{24}$&1.91&1.55&2.52&-& -9.20&	-10.84
	\\ \hline
	Cs$_8$Ag$_4$Ga$_{2}$Bi$_{2}$Cl$_{24}$&1.57&1.38&2.45&-& -8.98
	&-\\ \hline
	Cs$_8$Ag$_4$Ga$_{3}$Bi$_{1}$Cl$_{24}$& 2.17&2.03&3.11&-& -8.71&	-10.17
	\\ \hline
	Cs$_8$Ag$_4$Ga$_4$Cl$_{24}$&1.31&1.27&2.62&-&-8.33&-9.70\\ \hline
	Cs$_8$Ag$_4$In$_{1}$Bi$_{3}$Cl$_{24}$&1.91&1.61&2.58&-& -9.43&-\\ \hline
	Cs$_8$Ag$_4$In$_{2}$Bi$_{2}$Cl$_{24}$&1.52&1.37&2.40&-& -9.47&	-11.30\\ \hline
	Cs$_8$Ag$_4$In$_{3}$Bi$_{1}$Cl$_{24}$&2.04&1.97&3.07&-& -9.16&-
	\\ \hline
	Cs$_8$Ag$_4$In$_4$Cl$_{24}$&1.19&1.17&2.56&-&-9.46&-11.38\\ \hline
	Cs$_8$Ag$_4$Sb$_{1}$Bi$_{3}$Cl$_{24}$&1.80&1.44&2.31&-& -9.30&-10.94
	\\ \hline
	Cs$_8$Ag$_4$Sb$_{2}$Bi$_{2}$Cl$_{24}$&1.71&1.37&2.22&-& -9.23&-10.76
	\\ \hline
	Cs$_8$Ag$_4$Sb$_{3}$Bi$_{1}$Cl$_{24}$&1.68&1.34&2.20&-& -9.16&	-10.59\\ \hline
	Cs$_8$Ag$_4$Sb$_4$Cl$_{24}$&1.69&1.64&2.57&-& -9.09&-\\ \hline
	Cs$_8$Ag$_4$Sc$_{1}$Bi$_{3}$Cl$_{24}$&2.13&1.70&-&-& -9.27&-\\ \hline
	Cs$_8$Ag$_4$Sc$_{2}$Bi$_{2}$Cl$_{24}$&2.18&1.69&-&-& -9.15&-\\ \hline
	Cs$_8$Ag$_4$Sc$_{3}$Bi$_{1}$Cl$_{24}$&2.71&2.14&-&-& -9.01&-\\ \hline
	Cs$_8$Ag$_4$Sc$_{4}$Cl$_{24}$&3.32&3.29&-&-& -8.84&- \\ \hline
	Cs$_8$Ag$_4$Tl$_{1}$Bi$_{3}$Cl$_{24}$&0.65&0.62&-&-&-&-\\ \hline
	Cs$_8$Ag$_4$Tl$_{2}$Bi$_{2}$Cl$_{24}$&0.45&0.42&-&-&-&-\\ \hline
	Cs$_8$Ag$_4$Tl$_{3}$Bi$_{1}$Cl$_{24}$&0.83&0.82&0.45&0.42&-&-\\ \hline			
	Cs$_8$Ag$_4$Tl$_{4}$Cl$_{24}$&0.40&0.39&-&-&-&-\\ \hline
	Cs$_8$Ag$_4$Y$_{1}$Bi$_{3}$Cl$_{24}$&2.11&1.74&-&-& -9.28&-\\ \hline			
	Cs$_8$Ag$_4$Y$_{2}$Bi$_{2}$Cl$_{24}$&2.28&1.80&-&-&-9.19&-\\ \hline				
	Cs$_8$Ag$_4$Y$_{3}$Bi$_{1}$Cl$_{24}$&2.87&2.21&-&-& -9.11&-\\ \hline
	Cs$_8$Ag$_4$Y$_{4}$Cl$_{24}$&3.73&3.69&-&-&-9.01&-\\ \hline
	
\end{tabular}
\label{Table1}
\end{table}	
\newpage
\begin{table}
\caption{Band gap and enthalpy of decomposition of different conformers (double perovskites) for M(I) substitution} 
\centering
\begin{tabular}{p{3.8cm} p{1.2cm}p{1.2cm}p{1.2cm}p{1.2cm}p{2.2cm}p{2.4cm}} \hline
	Conformers & PBE (eV)& PBE +SOC (eV) & HSE06 +SOC (eV)& HSE06 (eV)& $\Delta$H$_{\textrm{D}}$ (PBE+SOC) (eV) &$\Delta$H$_{\textrm{D}}$ (HSE06+SOC) (eV)\\ \hline		
	
	Cs$_8$Ag$_4$Bi$_4$Cl$_{24}$ & 2.06 & 1.67 & 2.60 & 3.15& -9.36&	-11.11\\ \hline		
	Cs$_8$Ag$_{3}$Au$_{1}$Bi$_4$Cl$_{24}$ & 1.15 & 0.88 & 1.79 & 2.17& -9.18&	-10.90\\ \hline
	Cs$_8$Ag$_{2}$Au$_{2}$Bi$_4$Cl$_{24}$ & 1.02 & 0.76 & 1.62 & 2.01& -9.00&	-10.70\\ \hline
	Cs$_8$Ag$_{1}$Au$_{3}$Bi$_4$Cl$_{24}$ & 0.96 & 0.69 &  1.54 & 1.94& -8.80	&-10.49\\ \hline			
	Cs$_8$Au$_4$Bi$_4$Cl$_{24}$ & 1.46 & 0.70 & 1.54 & 1.95& -8.62&	-10.28 \\ \hline
	Cs$_8$Ag$_{3}$Cu$_{1}$Bi$_4$Cl$_{24}$ & 1.27 & 1.07 & 2.23 & 2.61& -9.26&	-10.97\\ \hline
	Cs$_8$Ag$_{2}$Cu$_{2}$Bi$_4$Cl$_{24}$ & 1.20 & 1.00 & 2.12 & 2.54 & -9.15&	-10.83\\ \hline
	Cs$_8$Ag$_{1}$Cu$_{3}$Bi$_4$Cl$_{24}$ & 1.15 & 0.94 & 2.01 & 2.49& -9.04&	-10.69 \\ \hline
	Cs$_8$Cu$_4$Bi$_4$Cl$_{24}$ & 1.13 & 0.89 & 1.95 & 2.48& -8.92	&-10.54
	\\ \hline
	Cs$_8$Ag$_{3}$In$_{1}$Bi$_4$Cl$_{24}$ & 0.45 & 0.07 & 0.39 & 1.22& -9.25&	-11.01\\ \hline
	Cs$_8$Ag$_{2}$In$_{2}$Bi$_4$Cl$_{24}$ & 0.30 & 0.13 & 0.11 &1.10& -9.14&	-10.90\\ \hline
	Cs$_8$Ag$_{1}$In$_{3}$Bi$_4$Cl$_{24}$ & 0.14 & 0.23 & 0.17 & 0.98& -9.02&	-10.80\\ \hline
	Cs$_8$In$_4$Bi$_4$Cl$_{24}$ & 0.29 & 0.55 & 0.46 & -& -8.89&-10.71\\ \hline
	Cs$_8$Ag$_{3}$K$_{1}$Bi$_4$Cl$_{24}$ &2.07&1.71&-&-& -10.37&-\\ \hline
	Cs$_8$Ag$_{2}$K$_{2}$Bi$_4$Cl$_{24}$ & 2.15&1.80&-&-& -10.37&-\\ \hline
	Cs$_8$Ag$_{1}$K$_{3}$Bi$_4$Cl$_{24}$ &2.32&1.97&-&-& -10.86&-\\ \hline
	Cs$_8$K$_4$Bi$_4$Cl$_{24}$ & 4.18 & 3.31&-&-& -11.34&-\\ \hline
	Cs$_8$Ag$_{3}$Na$_{1}$Bi$_4$Cl$_{24}$ &2.06 & 1.69&-&-& -9.36&-\\ \hline
	Cs$_8$Ag$_{2}$Na$_{2}$Bi$_4$Cl$_{24}$ &2.16&1.78&-&-& -9.36&-\\ \hline
	Cs$_8$Ag$_{1}$Na$_{3}$Bi$_4$Cl$_{24}$&2.35&1.96&-&-& -9.36&-\\ \hline
	Cs$_8$Na$_4$Bi$_4$Cl$_{24}$& 3.12& 3.09&-&-& -9.35 &-\\ \hline
	Cs$_8$Ag$_{3}$Ti$_{1}$Bi$_4$Cl$_{24}$&0.20&0.02&0.05&0.40 & -9.10&	-11.04 \\ \hline
	Cs$_8$Ag$_{2}$Ti$_{2}$Bi$_4$Cl$_{24}$&0.03&0.01&-&-& -8.88&	-11.57
	\\ \hline
	Cs$_8$Ag$_{1}$Ti$_{3}$Bi$_4$Cl$_{24}$&0.05&0.01&0.06&-& -8.71& -11.91\\ \hline
	Cs$_8$Ti$_4$Bi$_4$Cl$_{24}$&0.04& 0.01&0.02&-& -8.56&-12.51\\ \hline		
\end{tabular}
\label{Table2}
\end{table}
\newpage
\begin{table}
\caption{Band gap and enthalpy of decomposition of different conformers (double perovskites) for M(II) substitution} 
\centering
\begin{tabular}{p{3.8cm} p{1.2cm}p{1.2cm}p{1.2cm}p{1.2cm}p{2.2cm}p{2.4cm}} \hline
	Conformers & PBE (eV)& PBE +SOC (eV) & HSE06 +SOC (eV)& HSE06 (eV)& $\Delta$H$_{\textrm{D}}$ (PBE+SOC) (eV) &$\Delta$H$_{\textrm{D}}$ (HSE06+SOC) (eV)\\ \hline			
	Cs$_8$Ag$_{3}$Cd$_{2}$Bi$_{3}$Cl$_{24}$&1.12&1.08&1.96&2.14& -9.12&	-10.76\\ \hline
	Cs$_8$Ag$_{3}$Co$_{2}$Bi$_{3}$Cl$_{24}$&0.77&0.71&2.01&-& -8.24&	-11.79 \\ \hline
	Cs$_8$Ag$_{3}$Cu$_{2}$Bi$_{3}$Cl$_{24}$&Metal&Metal&-&-& -8.87&-
	\\ \hline	
	Cs$_8$Ag$_{3}$Ge$_{2}$Bi$_{3}$Cl$_{24}$&1.06&0.56&1.27&2.12& -9.18&	-11.50\\ \hline
	Cs$_8$Ag$_{3}$Mn$_{2}$Bi$_{3}$Cl$_{24}$&1.00&0.77&2.02&2.31& -9.56&	-13.20
	\\ \hline
	Cs$_8$Ag$_{3}$Mo$_{2}$Bi$_{3}$Cl$_{24}$&Metal&Metal&-&-& -8.31&-
	\\ \hline
	Cs$_8$Ag$_{3}$Ni$_{2}$Bi$_{3}$Cl$_{24}$&0.79&0.07&1.64&1.73& -8.70	&-11.63\\ \hline
	Cs$_8$Ag$_{3}$Sn$_{2}$Bi$_{3}$Cl$_{24}$&1.05&0.41&1.06&2.07& -9.17&	-10.87\\ \hline
	Cs$_8$Ag$_{3}$V$_{2}$Bi$_{3}$Cl$_{24}$&1.90&0.31&0.77&1.26& -8.87&	-11.74\\ \hline
	Cs$_8$Ag$_{3}$Zn$_{2}$Bi$_{3}$Cl$_{24}$&1.04&0.93&1.87&2.05& -8.80&	-10.43\\ \hline
	Cs$_8$Ag$_{3}$Rh$_{2}$Bi$_{3}$Cl$_{24}$&0.50&0.19&1.32&1.76& -9.38&	-11.99\\ \hline
\end{tabular}
\label{Table3}
\end{table}	
\clearpage	
\section{{Band structures for Au and Sn substitution}}
\begin{figure}[H]
	\includegraphics[width=0.8\columnwidth,clip]{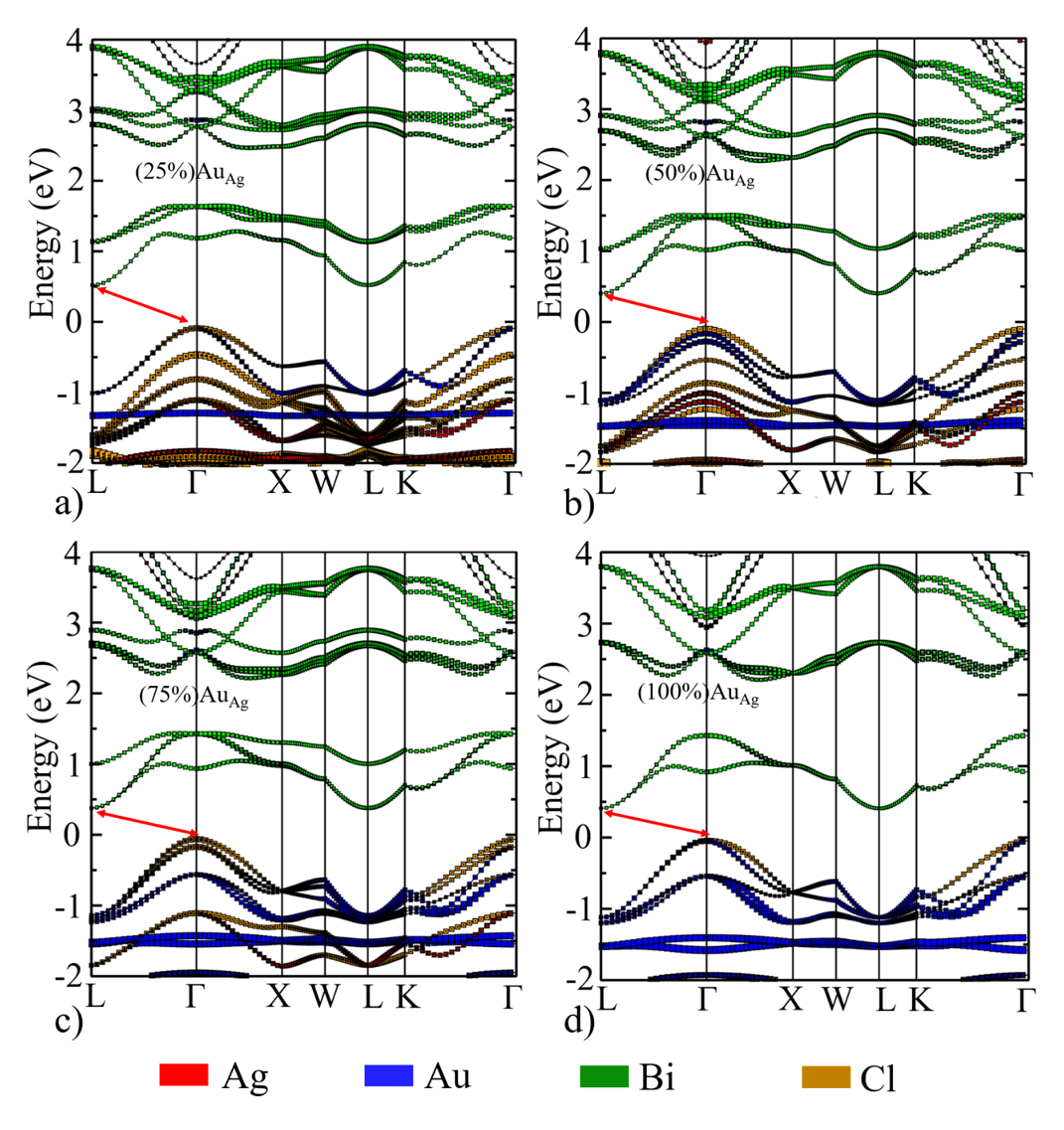}
	\caption{Band structure of (a) Cs$_8$Ag$_3$Au$_1$Bi$_4$Cl$_{24}$, (b) Cs$_8$Ag$_2$Au$_2$Bi$_4$Cl$_{24}$, (c) Cs$_8$Ag$_1$Au$_3$Bi$_4$Cl$_{24}$ and (d) Cs$_8$Au$_4$Bi$_4$Cl$_{24}$, using PBE+SOC $\epsilon_{xc}$ functional.}
	\label{fig4}
\end{figure}
\begin{figure}[H]
	\includegraphics[width=0.6\columnwidth,clip]{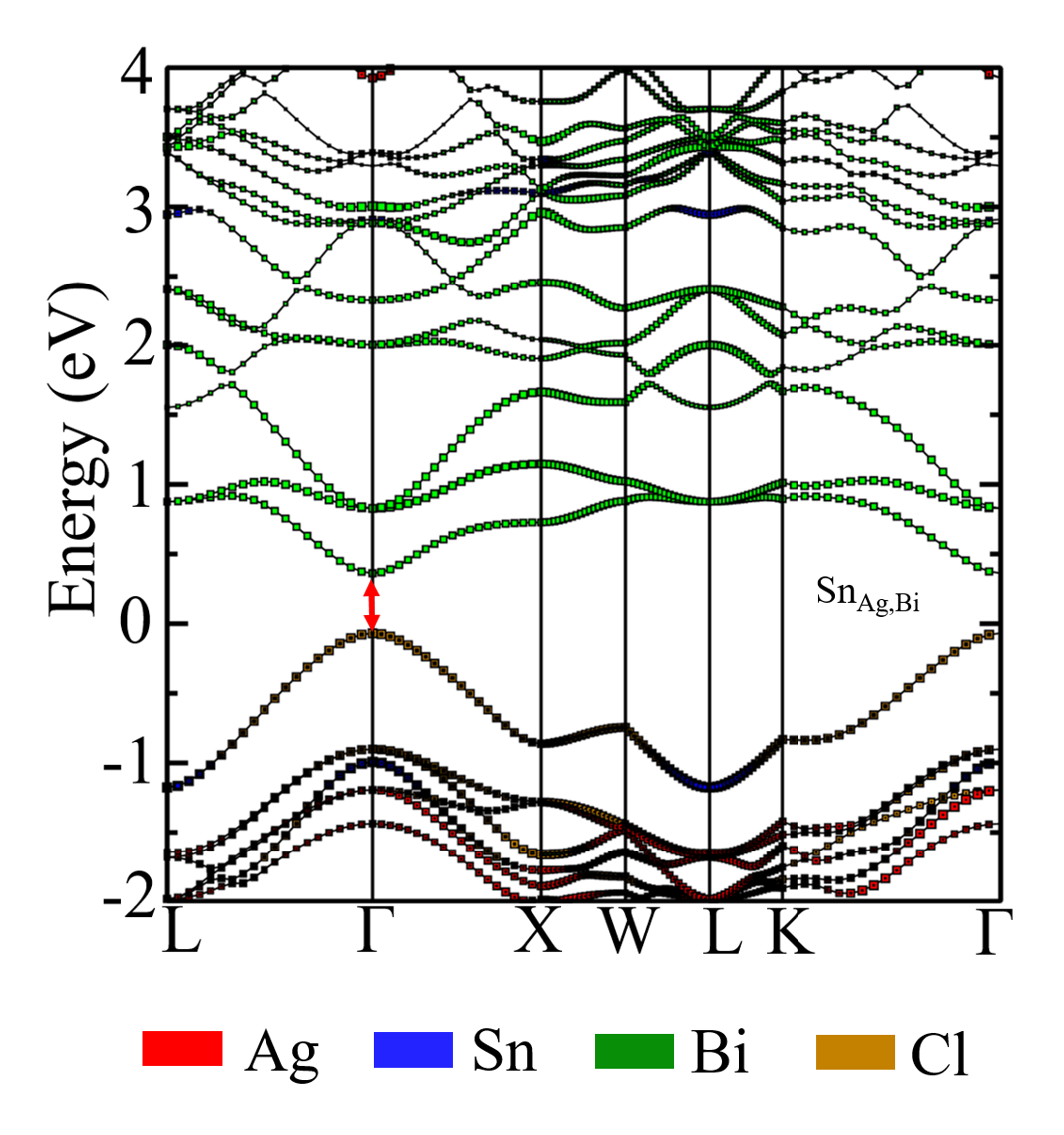}
	\caption{Band structure of Cs$_8$Ag$_3$Sn$_2$Bi$_3$Cl$_{24}$, using PBE+SOC $\epsilon_{xc}$ functional.}
	\label{fig5}
\end{figure}
\section{{Partial density of states (pDOS) plot, showing contribution of various orbitals in valence band maximum (VBM) and conduction band minimum (CBm) of few selected conformers}}
\begin{figure}[H]
	\centering
	\includegraphics[width=0.8\columnwidth,clip]{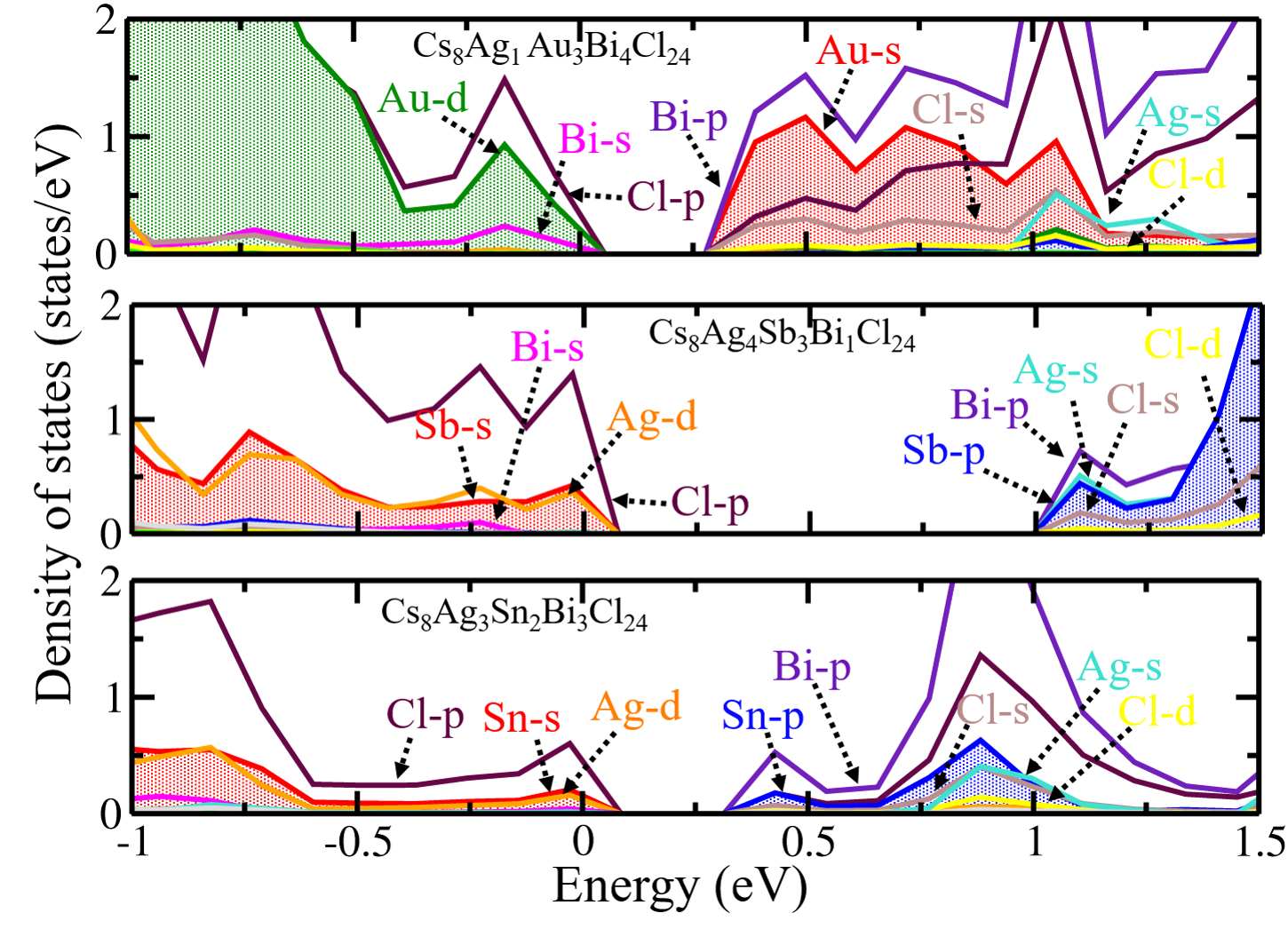}
	\caption{Partial density of states (pDOS) for Cs$_8$Ag$_{1}$Au$_{3}$Bi$_4$Cl$_{24}$, Cs$_8$Ag$_4$Sb$_{3}$Bi$_{1}$Cl$_{24}$ and Cs$_8$Ag$_{3}$Sn$_{2}$Bi$_{3}$Cl$_{24}$, using PBE+SOC $\epsilon_{xc}$ functional.}
	\label{fig6}
\end{figure}
In Figure \ref{fig6}, through partial density of states (pDOS), we have presented a clear picture of different orbitals' contribution in VBM and CBm. We have noticed that molecular orbitals that contribute to VBM and CBm consist of hybridization of different atomic orbitals. Hence, complete substitution of Bi with Sb, eliminates the contribution of Bi s-orbital in CBm. As a consequence, the band gap increases. Similarly, in some cases, for 100\% removal of Ag and Bi, a sudden change in band gap has been observed, which is ascribed to the elimination of non-degenerate energy levels. 
\section{{Optical properties using HSE06 $\epsilon_{xc}$ functional}}
\begin{figure}[H]
	\centering
	\includegraphics[width=0.8\columnwidth,clip]{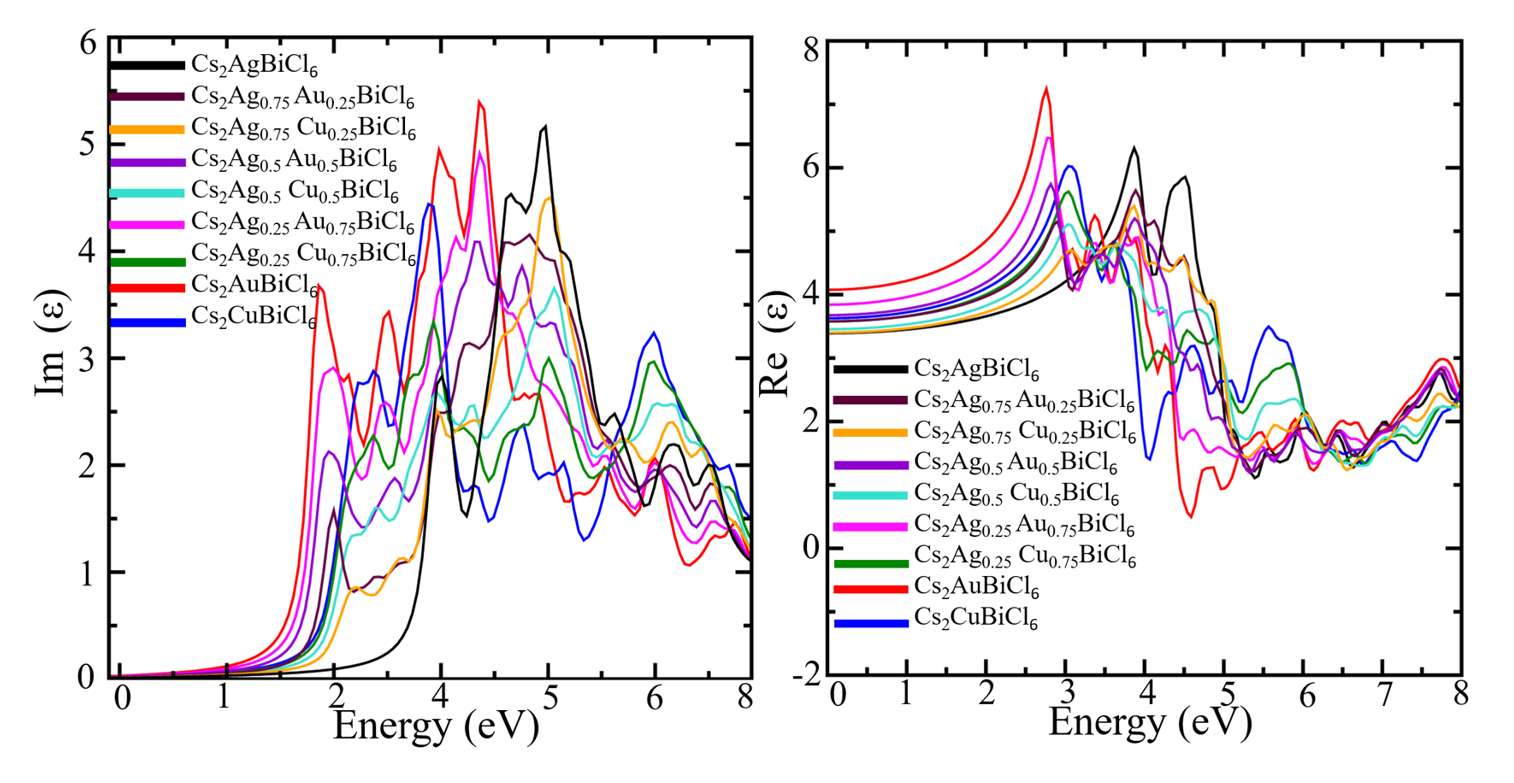}
	\caption{Variation of imaginary and real part of dielectric constant w.r.t. energy for double perovskites alloyed with monovalent cations, using HSE06 $\epsilon_{xc}$ functional.}
	\label{fig7}
\end{figure}

\begin{figure}[H]
	\centering
	\includegraphics[width=0.8\columnwidth,clip]{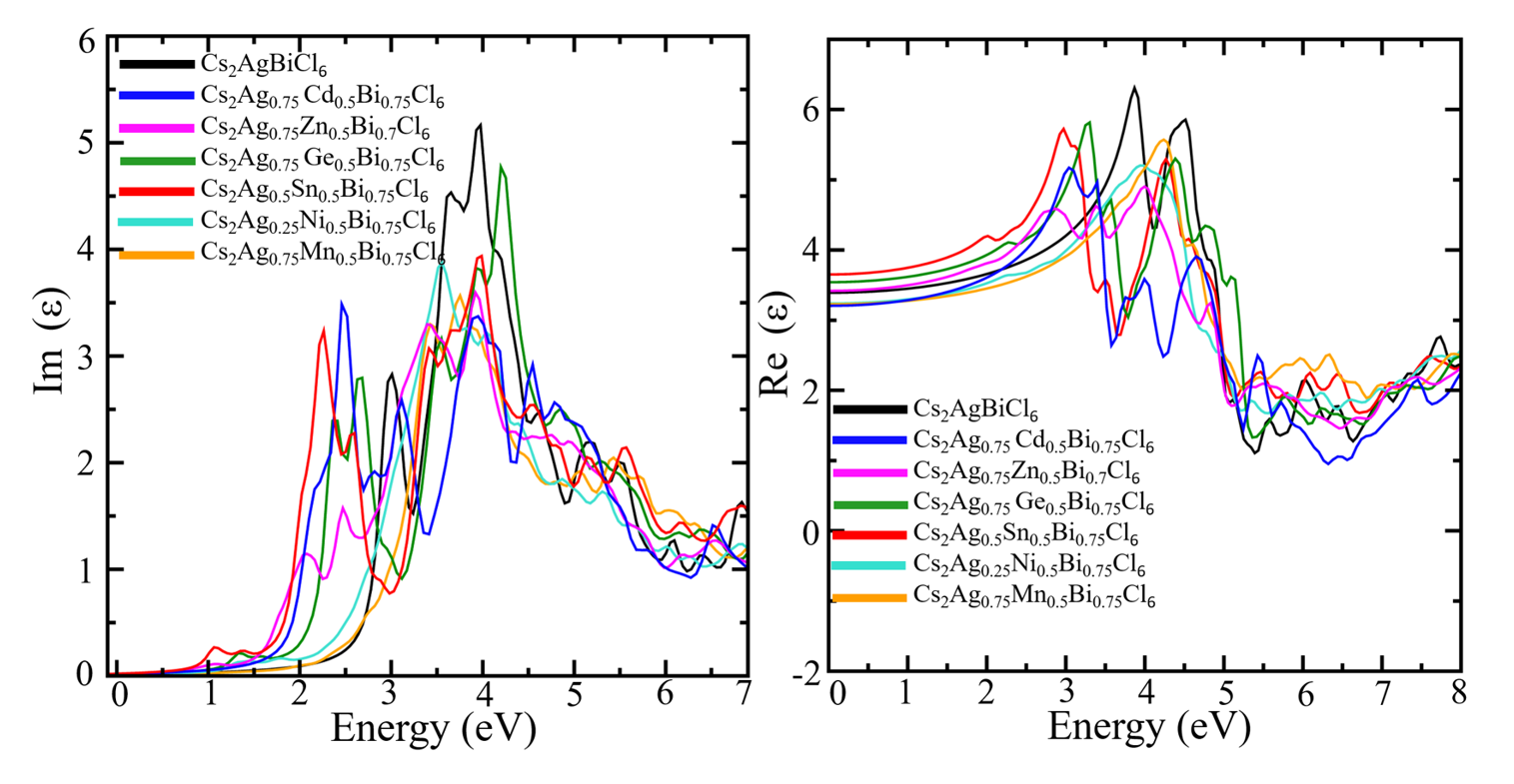}
	\caption{Variation of imaginary and real part of dielectric constant w.r.t. energy for double perovskites alloyed with divalent cations, using HSE06 $\epsilon_{xc}$ functional.}
	\label{fig8}
\end{figure}

\begin{mcitethebibliography}{49}
\providecommand*\natexlab[1]{#1}
\providecommand*\mciteSetBstSublistMode[1]{}
\providecommand*\mciteSetBstMaxWidthForm[2]{}
\providecommand*\mciteBstWouldAddEndPuncttrue
  {\def\EndOfBibitem{\unskip.}}
\providecommand*\mciteBstWouldAddEndPunctfalse
  {\let\EndOfBibitem\relax}
\providecommand*\mciteSetBstMidEndSepPunct[3]{}
\providecommand*\mciteSetBstSublistLabelBeginEnd[3]{}
\providecommand*\EndOfBibitem{}
\mciteSetBstSublistMode{f}
\mciteSetBstMaxWidthForm{subitem}{(\alph{mcitesubitemcount})}
\mciteSetBstSublistLabelBeginEnd
  {\mcitemaxwidthsubitemform\space}
  {\relax}
  {\relax}

\bibitem[Frost \latin{et~al.}(2014)Frost, Butler, Brivio, Hendon,
  Van~Schilfgaarde, and Walsh]{frost2014atomistic}
Frost,~J.~M.; Butler,~K.~T.; Brivio,~F.; Hendon,~C.~H.; Van~Schilfgaarde,~M.;
  Walsh,~A. Atomistic origins of high-performance in hybrid halide perovskite
  solar cells. \emph{Nano letters} \textbf{2014}, \emph{14}, 2584--2590\relax
\mciteBstWouldAddEndPuncttrue
\mciteSetBstMidEndSepPunct{\mcitedefaultmidpunct}
{\mcitedefaultendpunct}{\mcitedefaultseppunct}\relax
\EndOfBibitem
\bibitem[De~Marco \latin{et~al.}(2016)De~Marco, Zhou, Chen, Sun, Liu, Meng,
  Yao, Liu, Schiffer, and Yang]{de2016guanidinium}
De~Marco,~N.; Zhou,~H.; Chen,~Q.; Sun,~P.; Liu,~Z.; Meng,~L.; Yao,~E.-P.;
  Liu,~Y.; Schiffer,~A.; Yang,~Y. Guanidinium: a route to enhanced carrier
  lifetime and open-circuit voltage in hybrid perovskite solar cells.
  \emph{Nano letters} \textbf{2016}, \emph{16}, 1009--1016\relax
\mciteBstWouldAddEndPuncttrue
\mciteSetBstMidEndSepPunct{\mcitedefaultmidpunct}
{\mcitedefaultendpunct}{\mcitedefaultseppunct}\relax
\EndOfBibitem
\bibitem[Zhang \latin{et~al.}(2018)Zhang, Liang, Wang, Guo, Sun, and
  Xu]{zhang2018planar}
Zhang,~Y.; Liang,~Y.; Wang,~Y.; Guo,~F.; Sun,~L.; Xu,~D. Planar
  FAPb${\mathrm{Br}_{3}}$ Solar Cells with Power Conversion Efficiency above
  10\%. \emph{American Chemical Society Energy Letters} \textbf{2018},
  \emph{3}, 1808--1814\relax
\mciteBstWouldAddEndPuncttrue
\mciteSetBstMidEndSepPunct{\mcitedefaultmidpunct}
{\mcitedefaultendpunct}{\mcitedefaultseppunct}\relax
\EndOfBibitem
\bibitem[Kojima \latin{et~al.}(2009)Kojima, Teshima, and Shirai]{r1}
Kojima,~A.; Teshima,~K.; Shirai,~Y. Tsutomu Miyasaka Organo metal Halide
  Perovskites as Visible-Light Sensitizers for Photovoltaic Cells.
  \emph{Journal of the American Chemical Society} \textbf{2009}, \emph{131},
  6050--6051\relax
\mciteBstWouldAddEndPuncttrue
\mciteSetBstMidEndSepPunct{\mcitedefaultmidpunct}
{\mcitedefaultendpunct}{\mcitedefaultseppunct}\relax
\EndOfBibitem
\bibitem[Savenije \latin{et~al.}(2014)Savenije, Ponseca, Kunneman, Abdellah,
  Zheng, Tian, Zhu, Canton, Scheblykin, Pullerits, Yartsev, and
  Sundström]{doi:10.1021/jz500858a}
Savenije,~T.~J.; Ponseca,~C.~S.; Kunneman,~L.; Abdellah,~M.; Zheng,~K.;
  Tian,~Y.; Zhu,~Q.; Canton,~S.~E.; Scheblykin,~I.~G.; Pullerits,~T.
  \latin{et~al.}  Thermally Activated Exciton Dissociation and Recombination
  Control the Carrier Dynamics in Organometal Halide Perovskite. \emph{The
  Journal of Physical Chemistry Letters} \textbf{2014}, \emph{5},
  2189--2194\relax
\mciteBstWouldAddEndPuncttrue
\mciteSetBstMidEndSepPunct{\mcitedefaultmidpunct}
{\mcitedefaultendpunct}{\mcitedefaultseppunct}\relax
\EndOfBibitem
\bibitem[Basera \latin{et~al.}(2020)Basera, Kumar, Saini, and
  Bhattacharya]{PhysRevB.101.054108}
Basera,~P.; Kumar,~M.; Saini,~S.; Bhattacharya,~S. Reducing lead toxicity in
  the methylammonium lead halide ${\mathrm{MAPbI}}_{3}$: Why Sn substitution
  should be preferred to Pb vacancy for optimum solar cell efficiency.
  \emph{Physical Review B} \textbf{2020}, \emph{101}, 054108\relax
\mciteBstWouldAddEndPuncttrue
\mciteSetBstMidEndSepPunct{\mcitedefaultmidpunct}
{\mcitedefaultendpunct}{\mcitedefaultseppunct}\relax
\EndOfBibitem
\bibitem[Steirer \latin{et~al.}(2016)Steirer, Schulz, Teeter, Stevanovic, Yang,
  Zhu, and Berry]{doi:10.1021/acsenergylett.6b00196}
Steirer,~K.~X.; Schulz,~P.; Teeter,~G.; Stevanovic,~V.; Yang,~M.; Zhu,~K.;
  Berry,~J.~J. Defect Tolerance in Methylammonium Lead Triiodide Perovskite.
  \emph{ACS Energy Letters} \textbf{2016}, \emph{1}, 360--366\relax
\mciteBstWouldAddEndPuncttrue
\mciteSetBstMidEndSepPunct{\mcitedefaultmidpunct}
{\mcitedefaultendpunct}{\mcitedefaultseppunct}\relax
\EndOfBibitem
\bibitem[Yang \latin{et~al.}(2015)Yang, Noh, Jeon, Kim, Ryu, Seo, and Seok]{r2}
Yang,~W.~S.; Noh,~J.~H.; Jeon,~N.~J.; Kim,~Y.~C.; Ryu,~S.; Seo,~J.; Seok,~S.~I.
  High-performance photovoltaic perovskite layers fabricated through
  intramolecular exchange. \emph{Science} \textbf{2015}, \emph{348},
  1234--1237\relax
\mciteBstWouldAddEndPuncttrue
\mciteSetBstMidEndSepPunct{\mcitedefaultmidpunct}
{\mcitedefaultendpunct}{\mcitedefaultseppunct}\relax
\EndOfBibitem
\bibitem[Huang \latin{et~al.}(2016)Huang, Shi, Zhu, Li, Luo, and Meng]{r3}
Huang,~H.; Shi,~J.; Zhu,~L.; Li,~D.; Luo,~Y.; Meng,~Q. Two-step ultrasonic
  spray deposition of ${\mathrm{CH}}_{3}{\mathrm{NH}}_{3}{\mathrm{PbI}}_{3}$
  for efficient and large-area perovskite solar cell. \emph{Nano Energy}
  \textbf{2016}, \emph{27}, 352--358\relax
\mciteBstWouldAddEndPuncttrue
\mciteSetBstMidEndSepPunct{\mcitedefaultmidpunct}
{\mcitedefaultendpunct}{\mcitedefaultseppunct}\relax
\EndOfBibitem
\bibitem[Park and Seok(2019)Park, and Seok]{park2019intrinsic}
Park,~B.-w.; Seok,~S.~I. Intrinsic instability of inorganic--organic hybrid
  halide perovskite materials. \emph{Advanced Materials} \textbf{2019},
  \emph{31}, 1805337\relax
\mciteBstWouldAddEndPuncttrue
\mciteSetBstMidEndSepPunct{\mcitedefaultmidpunct}
{\mcitedefaultendpunct}{\mcitedefaultseppunct}\relax
\EndOfBibitem
\bibitem[Babayigit \latin{et~al.}(2016)Babayigit, Ethirajan, Muller, and
  Conings]{babayigit2016toxicity}
Babayigit,~A.; Ethirajan,~A.; Muller,~M.; Conings,~B. Toxicity of organometal
  halide perovskite solar cells. \emph{Nature Materials} \textbf{2016},
  \emph{15}, 247\relax
\mciteBstWouldAddEndPuncttrue
\mciteSetBstMidEndSepPunct{\mcitedefaultmidpunct}
{\mcitedefaultendpunct}{\mcitedefaultseppunct}\relax
\EndOfBibitem
\bibitem[Kulbak \latin{et~al.}(2015)Kulbak, Cahen, and
  Hodes]{kulbak2015important}
Kulbak,~M.; Cahen,~D.; Hodes,~G. How important is the organic part of lead
  halide perovskite photovoltaic cells? Efficient ${\mathrm{CsPbBr}}_{3}$
  cells. \emph{The Journal of Physical Chemistry Letters} \textbf{2015},
  \emph{6}, 2452--2456\relax
\mciteBstWouldAddEndPuncttrue
\mciteSetBstMidEndSepPunct{\mcitedefaultmidpunct}
{\mcitedefaultendpunct}{\mcitedefaultseppunct}\relax
\EndOfBibitem
\bibitem[Liang \latin{et~al.}(2016)Liang, Wang, Wang, Xu, Lu, Ma, Zhu, Hu,
  Xiao, Yi, \latin{et~al.} others]{liang2016all}
Liang,~J.; Wang,~C.; Wang,~Y.; Xu,~Z.; Lu,~Z.; Ma,~Y.; Zhu,~H.; Hu,~Y.;
  Xiao,~C.; Yi,~X. \latin{et~al.}  All-inorganic perovskite solar cells.
  \emph{Journal of the American Chemical Society} \textbf{2016}, \emph{138},
  15829--15832\relax
\mciteBstWouldAddEndPuncttrue
\mciteSetBstMidEndSepPunct{\mcitedefaultmidpunct}
{\mcitedefaultendpunct}{\mcitedefaultseppunct}\relax
\EndOfBibitem
\bibitem[Noel \latin{et~al.}(2014)Noel, Stranks, Abate, Wehrenfennig, Guarnera,
  Haghighirad, Sadhanala, Eperon, Pathak, Johnston, \latin{et~al.} others]{S1}
Noel,~N.~K.; Stranks,~S.~D.; Abate,~A.; Wehrenfennig,~C.; Guarnera,~S.;
  Haghighirad,~A.-A.; Sadhanala,~A.; Eperon,~G.~E.; Pathak,~S.~K.;
  Johnston,~M.~B. \latin{et~al.}  Lead-free organic--inorganic tin halide
  perovskites for photovoltaic applications. \emph{Energy \& Environmental
  Science} \textbf{2014}, \emph{7}, 3061--3068\relax
\mciteBstWouldAddEndPuncttrue
\mciteSetBstMidEndSepPunct{\mcitedefaultmidpunct}
{\mcitedefaultendpunct}{\mcitedefaultseppunct}\relax
\EndOfBibitem
\bibitem[Hao \latin{et~al.}(2014)Hao, Stoumpos, Cao, Chang, and Kanatzidis]{S2}
Hao,~F.; Stoumpos,~C.~C.; Cao,~D.~H.; Chang,~R.~P.; Kanatzidis,~M.~G. Lead-free
  solid-state organic--inorganic halide perovskite solar cells. \emph{Nature
  Photonics} \textbf{2014}, \emph{8}, 489\relax
\mciteBstWouldAddEndPuncttrue
\mciteSetBstMidEndSepPunct{\mcitedefaultmidpunct}
{\mcitedefaultendpunct}{\mcitedefaultseppunct}\relax
\EndOfBibitem
\bibitem[Filip and Giustino(2016)Filip, and Giustino]{S3}
Filip,~M.~R.; Giustino,~F. Computational screening of homovalent lead
  substitution in organic--inorganic halide perovskites. \emph{The Journal of
  Physical Chemistry C} \textbf{2016}, \emph{120}, 166--173\relax
\mciteBstWouldAddEndPuncttrue
\mciteSetBstMidEndSepPunct{\mcitedefaultmidpunct}
{\mcitedefaultendpunct}{\mcitedefaultseppunct}\relax
\EndOfBibitem
\bibitem[K{\"o}rbel \latin{et~al.}(2016)K{\"o}rbel, Marques, and Botti]{S4}
K{\"o}rbel,~S.; Marques,~M.~A.; Botti,~S. Stability and electronic properties
  of new inorganic perovskites from high-throughput ab initio calculations.
  \emph{Journal of Materials Chemistry C} \textbf{2016}, \emph{4},
  3157--3167\relax
\mciteBstWouldAddEndPuncttrue
\mciteSetBstMidEndSepPunct{\mcitedefaultmidpunct}
{\mcitedefaultendpunct}{\mcitedefaultseppunct}\relax
\EndOfBibitem
\bibitem[Cai \latin{et~al.}(2019)Cai, Xie, Teng, Harikesh, Ghosh, Huck,
  Persson, Mathews, Mhaisalkar, Sherburne, \latin{et~al.} others]{cai2019high}
Cai,~Y.; Xie,~W.; Teng,~Y.~T.; Harikesh,~P.; Ghosh,~B.; Huck,~P.;
  Persson,~K.~A.; Mathews,~N.; Mhaisalkar,~S.~G.; Sherburne,~M. \latin{et~al.}
  High-throughput Computational Study of Halide Double Perovskite Inorganic
  Compounds. \emph{Chemistry of Materials} \textbf{2019}, \emph{31},
  5392--5401\relax
\mciteBstWouldAddEndPuncttrue
\mciteSetBstMidEndSepPunct{\mcitedefaultmidpunct}
{\mcitedefaultendpunct}{\mcitedefaultseppunct}\relax
\EndOfBibitem
\bibitem[Chen \latin{et~al.}(2019)Chen, Cai, Li, Hills-Kimball, Yang, Que,
  Nagaoka, Liu, Yang, Dong, \latin{et~al.} others]{chen2019yb}
Chen,~N.; Cai,~T.; Li,~W.; Hills-Kimball,~K.; Yang,~H.; Que,~M.; Nagaoka,~Y.;
  Liu,~Z.; Yang,~D.; Dong,~A. \latin{et~al.}  Yb-and Mn-Doped Lead-Free Double
  Perovskite ${\mathrm{Cs}}_{2}{\mathrm{AgBiX}}_{6}$ (X= Cl--, Br--)
  Nanocrystals. \emph{American Chemical Society Applied Materials \&
  Interfaces} \textbf{2019}, \emph{11}, 16855--16863\relax
\mciteBstWouldAddEndPuncttrue
\mciteSetBstMidEndSepPunct{\mcitedefaultmidpunct}
{\mcitedefaultendpunct}{\mcitedefaultseppunct}\relax
\EndOfBibitem
\bibitem[Slavney \latin{et~al.}(2016)Slavney, Hu, Lindenberg, and
  Karunadasa]{S5}
Slavney,~A.~H.; Hu,~T.; Lindenberg,~A.~M.; Karunadasa,~H.~I. A bismuth-halide
  double perovskite with long carrier recombination lifetime for photovoltaic
  applications. \emph{Journal of the American Chemical Society} \textbf{2016},
  \emph{138}, 2138--2141\relax
\mciteBstWouldAddEndPuncttrue
\mciteSetBstMidEndSepPunct{\mcitedefaultmidpunct}
{\mcitedefaultendpunct}{\mcitedefaultseppunct}\relax
\EndOfBibitem
\bibitem[Volonakis \latin{et~al.}(2016)Volonakis, Filip, Haghighirad, Sakai,
  Wenger, Snaith, and Giustino]{S6}
Volonakis,~G.; Filip,~M.~R.; Haghighirad,~A.~A.; Sakai,~N.; Wenger,~B.;
  Snaith,~H.~J.; Giustino,~F. Lead-free halide double perovskites via
  heterovalent substitution of noble metals. \emph{The Journal of Physical
  Chemistry Letters} \textbf{2016}, \emph{7}, 1254--1259\relax
\mciteBstWouldAddEndPuncttrue
\mciteSetBstMidEndSepPunct{\mcitedefaultmidpunct}
{\mcitedefaultendpunct}{\mcitedefaultseppunct}\relax
\EndOfBibitem
\bibitem[Lamba \latin{et~al.}(2019)Lamba, Basera, Bhattacharya, and Sapra]{S7}
Lamba,~R.~S.; Basera,~P.; Bhattacharya,~S.; Sapra,~S. Band Gap Engineering in
  ${\mathrm{Cs}}_{2}({\mathrm{Na}}_{x}{\mathrm{Ag}}_{1-x}){\mathrm{BiCl}}_{6}$
  Double Perovskite Nanocrystals. \emph{The Journal of Physical Chemistry
  Letters} \textbf{2019}, \emph{10}, 5173--5181\relax
\mciteBstWouldAddEndPuncttrue
\mciteSetBstMidEndSepPunct{\mcitedefaultmidpunct}
{\mcitedefaultendpunct}{\mcitedefaultseppunct}\relax
\EndOfBibitem
\bibitem[Tran \latin{et~al.}(2017)Tran, Panella, Chamorro, Morey, and
  McQueen]{S8}
Tran,~T.~T.; Panella,~J.~R.; Chamorro,~J.~R.; Morey,~J.~R.; McQueen,~T.~M.
  Designing indirect--direct bandgap transitions in double perovskites.
  \emph{Materials Horizons} \textbf{2017}, \emph{4}, 688--693\relax
\mciteBstWouldAddEndPuncttrue
\mciteSetBstMidEndSepPunct{\mcitedefaultmidpunct}
{\mcitedefaultendpunct}{\mcitedefaultseppunct}\relax
\EndOfBibitem
\bibitem[Deng \latin{et~al.}(2017)Deng, Deng, He, Wang, Chen, Wei, and
  Feng]{S9}
Deng,~W.; Deng,~Z.-Y.; He,~J.; Wang,~M.; Chen,~Z.-X.; Wei,~S.-H.; Feng,~H.-J.
  Synthesis of ${\mathrm{Cs}}_{2}{\mathrm{AgSbCl}}_{6}$ and improved
  optoelectronic properties of
  ${\mathrm{Cs}}_{2}{\mathrm{AgSbCl}}_{6}$/${\mathrm{TiO}}_{2}$ heterostructure
  driven by the interface effect for lead-free double perovskites solar cells.
  \emph{Applied Physics Letters} \textbf{2017}, \emph{111}, 151602\relax
\mciteBstWouldAddEndPuncttrue
\mciteSetBstMidEndSepPunct{\mcitedefaultmidpunct}
{\mcitedefaultendpunct}{\mcitedefaultseppunct}\relax
\EndOfBibitem
\bibitem[Volonakis \latin{et~al.}(2017)Volonakis, Haghighirad, Milot, Sio,
  Filip, Wenger, Johnston, Herz, Snaith, and Giustino]{S10}
Volonakis,~G.; Haghighirad,~A.~A.; Milot,~R.~L.; Sio,~W.~H.; Filip,~M.~R.;
  Wenger,~B.; Johnston,~M.~B.; Herz,~L.~M.; Snaith,~H.~J.; Giustino,~F.
  ${\mathrm{Cs}}_{2}{\mathrm{InAgCl}}_{6}$: a new lead-free halide double
  perovskite with direct band gap. \emph{The Journal of Physical Chemistry
  Letters} \textbf{2017}, \emph{8}, 772--778\relax
\mciteBstWouldAddEndPuncttrue
\mciteSetBstMidEndSepPunct{\mcitedefaultmidpunct}
{\mcitedefaultendpunct}{\mcitedefaultseppunct}\relax
\EndOfBibitem
\bibitem[Locardi \latin{et~al.}(2018)Locardi, Cirignano, Baranov, Dang, Prato,
  Drago, Ferretti, Pinchetti, Fanciulli, Brovelli, \latin{et~al.}
  others]{locardi2018colloidal}
Locardi,~F.; Cirignano,~M.; Baranov,~D.; Dang,~Z.; Prato,~M.; Drago,~F.;
  Ferretti,~M.; Pinchetti,~V.; Fanciulli,~M.; Brovelli,~S. \latin{et~al.}
  Colloidal synthesis of double perovskite
  ${\mathrm{Cs}}_{2}{\mathrm{AgInCl}}_{6}$ and Mn-doped
  ${\mathrm{Cs}}_{2}{\mathrm{AgInCl}}_{6}$ nanocrystals. \emph{Journal of the
  American Chemical Society} \textbf{2018}, \emph{140}, 12989--12995\relax
\mciteBstWouldAddEndPuncttrue
\mciteSetBstMidEndSepPunct{\mcitedefaultmidpunct}
{\mcitedefaultendpunct}{\mcitedefaultseppunct}\relax
\EndOfBibitem
\bibitem[Zhao \latin{et~al.}(2017)Zhao, Yang, Fu, Yang, Xu, Yu, Wei, and
  Zhang]{S11}
Zhao,~X.-G.; Yang,~J.-H.; Fu,~Y.; Yang,~D.; Xu,~Q.; Yu,~L.; Wei,~S.-H.;
  Zhang,~L. Design of lead-free inorganic halide perovskites for solar cells
  via cation-transmutation. \emph{Journal of the American Chemical Society}
  \textbf{2017}, \emph{139}, 2630--2638\relax
\mciteBstWouldAddEndPuncttrue
\mciteSetBstMidEndSepPunct{\mcitedefaultmidpunct}
{\mcitedefaultendpunct}{\mcitedefaultseppunct}\relax
\EndOfBibitem
\bibitem[Meng \latin{et~al.}(2017)Meng, Wang, Xiao, Wang, Mitzi, and Yan]{S13}
Meng,~W.; Wang,~X.; Xiao,~Z.; Wang,~J.; Mitzi,~D.~B.; Yan,~Y. Parity-forbidden
  transitions and their impact on the optical absorption properties of
  lead-free metal halide perovskites and double perovskites. \emph{The Journal
  of Physical Chemistry Letters} \textbf{2017}, \emph{8}, 2999--3007\relax
\mciteBstWouldAddEndPuncttrue
\mciteSetBstMidEndSepPunct{\mcitedefaultmidpunct}
{\mcitedefaultendpunct}{\mcitedefaultseppunct}\relax
\EndOfBibitem
\bibitem[Perdew \latin{et~al.}(1992)Perdew, Chevary, Vosko, Jackson, Pederson,
  Singh, and Fiolhais]{perdew1992atoms}
Perdew,~J.~P.; Chevary,~J.~A.; Vosko,~S.~H.; Jackson,~K.~A.; Pederson,~M.~R.;
  Singh,~D.~J.; Fiolhais,~C. Atoms, molecules, solids, and surfaces:
  Applications of the generalized gradient approximation for exchange and
  correlation. \emph{Physical Review B} \textbf{1992}, \emph{46}, 6671\relax
\mciteBstWouldAddEndPuncttrue
\mciteSetBstMidEndSepPunct{\mcitedefaultmidpunct}
{\mcitedefaultendpunct}{\mcitedefaultseppunct}\relax
\EndOfBibitem
\bibitem[Heyd \latin{et~al.}(2003)Heyd, Scuseria, and
  Ernzerhof]{heyd2003hybrid}
Heyd,~J.; Scuseria,~G.~E.; Ernzerhof,~M. Hybrid functionals based on a screened
  Coulomb potential. \emph{The Journal of Chemical Physics} \textbf{2003},
  \emph{118}, 8207--8215\relax
\mciteBstWouldAddEndPuncttrue
\mciteSetBstMidEndSepPunct{\mcitedefaultmidpunct}
{\mcitedefaultendpunct}{\mcitedefaultseppunct}\relax
\EndOfBibitem
\bibitem[McClure \latin{et~al.}(2016)McClure, Ball, Windl, and
  Woodward]{mcclure2016cs2agbix6}
McClure,~E.~T.; Ball,~M.~R.; Windl,~W.; Woodward,~P.~M.
  ${\mathrm{Cs}}_{2}{\mathrm{AgBiX}}_{6}$ (X= Br, Cl): new visible light
  absorbing, lead-free halide perovskite semiconductors. \emph{Chemistry of
  Materials} \textbf{2016}, \emph{28}, 1348--1354\relax
\mciteBstWouldAddEndPuncttrue
\mciteSetBstMidEndSepPunct{\mcitedefaultmidpunct}
{\mcitedefaultendpunct}{\mcitedefaultseppunct}\relax
\EndOfBibitem
\bibitem[Kieslich \latin{et~al.}(2015)Kieslich, Sun, and
  Cheetham]{kieslich2015extended}
Kieslich,~G.; Sun,~S.; Cheetham,~A.~K. An extended tolerance factor approach
  for organic--inorganic perovskites. \emph{Chemical science} \textbf{2015},
  \emph{6}, 3430--3433\relax
\mciteBstWouldAddEndPuncttrue
\mciteSetBstMidEndSepPunct{\mcitedefaultmidpunct}
{\mcitedefaultendpunct}{\mcitedefaultseppunct}\relax
\EndOfBibitem
\bibitem[Li \latin{et~al.}(2008)Li, Lu, Ding, Feng, Gao, and
  Guo]{li2008formability}
Li,~C.; Lu,~X.; Ding,~W.; Feng,~L.; Gao,~Y.; Guo,~Z. Formability of
  ${\mathrm{ABX}}_{3}$ (X= F, Cl, Br, I) Halide Perovskites. \emph{Acta
  Crystallographica Section B: Structural Science} \textbf{2008}, \emph{64},
  702--707\relax
\mciteBstWouldAddEndPuncttrue
\mciteSetBstMidEndSepPunct{\mcitedefaultmidpunct}
{\mcitedefaultendpunct}{\mcitedefaultseppunct}\relax
\EndOfBibitem
\bibitem[Kangsabanik \latin{et~al.}(2018)Kangsabanik, Sugathan, Yadav, Yella,
  and Alam]{kangsabanik2018double}
Kangsabanik,~J.; Sugathan,~V.; Yadav,~A.; Yella,~A.; Alam,~A. Double
  perovskites overtaking the single perovskites: A set of new solar harvesting
  materials with much higher stability and efficiency. \emph{Physical Review
  Materials} \textbf{2018}, \emph{2}, 055401\relax
\mciteBstWouldAddEndPuncttrue
\mciteSetBstMidEndSepPunct{\mcitedefaultmidpunct}
{\mcitedefaultendpunct}{\mcitedefaultseppunct}\relax
\EndOfBibitem
\bibitem[Sun \latin{et~al.}(2018)Sun, Chen, and Yin]{S14}
Sun,~Q.; Chen,~H.; Yin,~W.-J. Do chalcogenide double perovskites work as solar
  cell absorbers: a first-principles study. \emph{Chemistry of Materials}
  \textbf{2018}, \emph{31}, 244--250\relax
\mciteBstWouldAddEndPuncttrue
\mciteSetBstMidEndSepPunct{\mcitedefaultmidpunct}
{\mcitedefaultendpunct}{\mcitedefaultseppunct}\relax
\EndOfBibitem
\bibitem[Shockley and Queisser(1961)Shockley, and
  Queisser]{shockley1961detailed}
Shockley,~W.; Queisser,~H.~J. Detailed balance limit of efficiency of p-n
  junction solar cells. \emph{Journal of Applied Physics} \textbf{1961},
  \emph{32}, 510--519\relax
\mciteBstWouldAddEndPuncttrue
\mciteSetBstMidEndSepPunct{\mcitedefaultmidpunct}
{\mcitedefaultendpunct}{\mcitedefaultseppunct}\relax
\EndOfBibitem
\bibitem[Kumar \latin{et~al.}(2020)Kumar, Jain, Singh, and Bhattacharya]{MK3}
Kumar,~M.; Jain,~M.; Singh,~A.; Bhattacharya,~S. Band gap Engineering by
  Sublattice Mixing in ${\mathrm{Cs}}_{2}{\mathrm{AgInCl}}_{6}$:
  High-throughput Screening from First-principles. \emph{arXiv preprint
  arXiv:2004.07991} \textbf{2020}, \relax
\mciteBstWouldAddEndPunctfalse
\mciteSetBstMidEndSepPunct{\mcitedefaultmidpunct}
{}{\mcitedefaultseppunct}\relax
\EndOfBibitem
\bibitem[Savory \latin{et~al.}(2016)Savory, Walsh, and Scanlon]{savory2016can}
Savory,~C.~N.; Walsh,~A.; Scanlon,~D.~O. Can Pb-free halide double perovskites
  support high-efficiency solar cells? \emph{American Chemical Society Energy
  Letters} \textbf{2016}, \emph{1}, 949--955\relax
\mciteBstWouldAddEndPuncttrue
\mciteSetBstMidEndSepPunct{\mcitedefaultmidpunct}
{\mcitedefaultendpunct}{\mcitedefaultseppunct}\relax
\EndOfBibitem
\bibitem[Basera \latin{et~al.}(2019)Basera, Saini, and
  Bhattacharya]{basera2019self}
Basera,~P.; Saini,~S.; Bhattacharya,~S. Self energy and excitonic effect in
  (un)doped ${\mathrm{TiO}}_{2}$ anatase: a comparative study of hybrid DFT, GW
  and BSE to explore optical properties. \emph{Journal of Materials Chemistry
  C} \textbf{2019}, \emph{7}, 14284--14293\relax
\mciteBstWouldAddEndPuncttrue
\mciteSetBstMidEndSepPunct{\mcitedefaultmidpunct}
{\mcitedefaultendpunct}{\mcitedefaultseppunct}\relax
\EndOfBibitem
\bibitem[Yu and Zunger(2012)Yu, and Zunger]{yu2012identification}
Yu,~L.; Zunger,~A. Identification of potential photovoltaic absorbers based on
  first-principles spectroscopic screening of materials. \emph{Physical Review
  Letters} \textbf{2012}, \emph{108}, 068701\relax
\mciteBstWouldAddEndPuncttrue
\mciteSetBstMidEndSepPunct{\mcitedefaultmidpunct}
{\mcitedefaultendpunct}{\mcitedefaultseppunct}\relax
\EndOfBibitem
\bibitem[Jain and Bhattacharya(2020, to be published)Jain, and
  Bhattacharya]{kk}
Jain,~M.; Bhattacharya,~S. Understanding the role of Sn substitution and Pb
  vacancy in ${\mathrm{FAPbBr}}_{3}$ perovskites: A hybrid functional study.
  \textbf{2020, to be published}, \relax
\mciteBstWouldAddEndPunctfalse
\mciteSetBstMidEndSepPunct{\mcitedefaultmidpunct}
{}{\mcitedefaultseppunct}\relax
\EndOfBibitem
\bibitem[Filip \latin{et~al.}(2016)Filip, Hillman, Haghighirad, Snaith, and
  Giustino]{filip2016band}
Filip,~M.~R.; Hillman,~S.; Haghighirad,~A.~A.; Snaith,~H.~J.; Giustino,~F. Band
  gaps of the lead-free halide double perovskites
  ${\mathrm{Cs}}_{2}{\mathrm{BiAgCl}}_{6}$ and
  ${\mathrm{Cs}}_{2}{\mathrm{BiAgBr}}_{6}$ from theory and experiment.
  \emph{The Journal of Physical Chemistry Letters} \textbf{2016}, \emph{7},
  2579--2585\relax
\mciteBstWouldAddEndPuncttrue
\mciteSetBstMidEndSepPunct{\mcitedefaultmidpunct}
{\mcitedefaultendpunct}{\mcitedefaultseppunct}\relax
\EndOfBibitem
\bibitem[Hohenberg and Kohn(1964)Hohenberg, and
  Kohn]{hohenberg1964inhomogeneous}
Hohenberg,~P.; Kohn,~W. Inhomogeneous electron gas. \emph{Physical Review}
  \textbf{1964}, \emph{136}, B864\relax
\mciteBstWouldAddEndPuncttrue
\mciteSetBstMidEndSepPunct{\mcitedefaultmidpunct}
{\mcitedefaultendpunct}{\mcitedefaultseppunct}\relax
\EndOfBibitem
\bibitem[Kohn and Sham(1965)Kohn, and Sham]{kohn1965self}
Kohn,~W.; Sham,~L.~J. Self-consistent equations including exchange and
  correlation effects. \emph{Physical Review} \textbf{1965}, \emph{140},
  A1133\relax
\mciteBstWouldAddEndPuncttrue
\mciteSetBstMidEndSepPunct{\mcitedefaultmidpunct}
{\mcitedefaultendpunct}{\mcitedefaultseppunct}\relax
\EndOfBibitem
\bibitem[Bl{\"o}chl(1994)]{blochl1994projector}
Bl{\"o}chl,~P.~E. Projector augmented-wave method. \emph{Physical Review B}
  \textbf{1994}, \emph{50}, 17953\relax
\mciteBstWouldAddEndPuncttrue
\mciteSetBstMidEndSepPunct{\mcitedefaultmidpunct}
{\mcitedefaultendpunct}{\mcitedefaultseppunct}\relax
\EndOfBibitem
\bibitem[Kresse and Hafner(1996)Kresse, and Hafner]{kresse199614251}
Kresse,~G.; Hafner,~J. 14251); g. kresse, j. furthm{\"u}ller. \emph{Physical
  Review B} \textbf{1996}, \emph{54}, 11169\relax
\mciteBstWouldAddEndPuncttrue
\mciteSetBstMidEndSepPunct{\mcitedefaultmidpunct}
{\mcitedefaultendpunct}{\mcitedefaultseppunct}\relax
\EndOfBibitem
\bibitem[Pulay(1980)]{pulay1980convergence}
Pulay,~P. Convergence acceleration of iterative sequences. The case of SCF
  iteration. \emph{Chemical Physics Letters} \textbf{1980}, \emph{73},
  393--398\relax
\mciteBstWouldAddEndPuncttrue
\mciteSetBstMidEndSepPunct{\mcitedefaultmidpunct}
{\mcitedefaultendpunct}{\mcitedefaultseppunct}\relax
\EndOfBibitem
\bibitem[Monkhorst and Pack(1976)Monkhorst, and Pack]{monkhorst1976special}
Monkhorst,~H.~J.; Pack,~J.~D. Special points for Brillouin-zone integrations.
  \emph{Physical Review B} \textbf{1976}, \emph{13}, 5188\relax
\mciteBstWouldAddEndPuncttrue
\mciteSetBstMidEndSepPunct{\mcitedefaultmidpunct}
{\mcitedefaultendpunct}{\mcitedefaultseppunct}\relax
\EndOfBibitem
\end{mcitethebibliography}
\providecommand{\latin}[1]{#1}
\makeatletter
\providecommand{\doi}
  {\begingroup\let\do\@makeother\dospecials
  \catcode`\{=1 \catcode`\}=2 \doi@aux}
\providecommand{\doi@aux}[1]{\endgroup\texttt{#1}}
\makeatother
\providecommand*\mcitethebibliography{\thebibliography}
\csname @ifundefined\endcsname{endmcitethebibliography}
  {\let\endmcitethebibliography\endthebibliography}{}

\end{document}